\documentclass[aps,prc,twocolumn,floatfix,showpacs,amsmath,amssymb]{revtex4}
\usepackage{graphicx}
\usepackage{amssymb}
\usepackage{dcolumn}
\usepackage{color} 
\begin{document}
\title{{\itshape Ab initio} no-core full configuration calculations of
light nuclei*}

\author{P. Maris,$^1$  J. P. Vary,$^{1}$ and A. M. Shirokov$^{1,2}$}
\affiliation{
$^1$Department of Physics and Astronomy, Iowa State University, Ames, Iowa 50011, USA \\
$^2$Skobeltsyn Institute of Nuclear Physics, Moscow State University, Moscow, 119991 Russia}

\date{\today}

\begin{abstract}
We perform no-core full configuration calculations for 
a set of light nuclei including $^{16}$O 
with a realistic NN
interaction, JISP16.  We obtain ground state energies and their
uncertainties through exponential extrapolations that we demonstrate
are reliable in $^2$H, $^3$H and $^4$He test cases where fully
converged results are obtained directly.  We find that $^6$He, $^6$Li and $^8$He
are underbound by about $600$~keV, $560$~keV and $1.7$~MeV,
respectively. $^{12}$C is overbound by about $1.7$~MeV and $^{16}$O is
overbound by about $16$~MeV. The first excited $0^+$ states in
$^{12}$C and $^{16}$O are also evaluated but their uncertainties 
are significantly larger than the uncertainties for the ground states.

\end{abstract}
\maketitle
%

\section{Introduction and Motivation}

The rapid development of {\it ab initio} methods for solving finite
nuclei has opened a range of nuclear phenomena that can be evaluated
to high precision using realistic nucleon-nucleon (NN) and
three-nucleon (NNN) interactions.  Such advances define a path for
testing fundamental properties of the strong interaction such as their
origins from QCD via chiral effective field theory
\cite{Weinberg,ORK94,N3LO,Navratil07}. 
In addition, they prepare a foundation for nuclear reaction theory
with unprecedented predictive power.

Here we investigate the direct solution of the nuclear many-body
problem by diagonalization in a sufficiently large basis space that
converged binding energies are accessed --- either directly or by simple
extrapolation.  Our choice is a traditional harmonic oscillator (HO) basis
so there are two basis space parameters, the HO 
energy $\hbar\Omega$ and the many-body basis space cutoff $N_{max}$.  
$N_{max}$ is defined as the maximum number of total oscillator
quanta allowed in the many-body basis space above the minimum for
that nucleus.
We obtain convergence in this 2-dimensional  parameter
space ($\hbar\Omega$,$N_{max}$), where convergence is defined as independence of both parameters within evaluated uncertainties.

Since we treat all nucleons equivalently and we
achieve convergence within evaluated uncertainties, we refer to our approach as the 
``No-Core Full Configuration'' (NCFC) method.  The NCFC is both related to and distinct from
the No-Core Shell Model (NCSM) \cite{NCSMC12}, that features a finite matrix truncation, 
and an effective Hamiltonian renormalized to that finite space. 
The regulator, $N_{max}$,  appears in our NCFC, where it is
taken to infinity,  and in the NCSM, where it also appears
in the definition of the effective Hamiltonian.  In both NCFC and NCSM, 
this choice of many-body 
basis regulator, $N_{max}$, is needed to preserve Galilean invariance --- to
factorize all solutions into a product of intrinsic and center-of-mass motion
components.  With $N_{max}$ as the regulator,  
both the NCFC and the NCSM  are distinguished 
from the ``Full Configuration Interaction" (FCI) method in atomic and molecular
physics that employs a cutoff in single-particle-space.

The NCFC results should agree with the NCSM and no-core FCI results when
the latter results are obtained in sufficiently large basis spaces.  In the 
case of NCSM, larger cluster sizes for the effective Hamiltonian 
may be employed to accelerate convergence.

Given the rapid advances in algorithms and computers, as well as the
development of realistic non-local NN interactions that facilitate
convergence, we are able to achieve converged results, either directly
or through extrapolation, without recourse to renormalization of the
interaction.  That is, with our adopted interaction, we do not need to
soften the NN interaction by treating it with an effective interaction
formalism.  Renormalization formalisms necessarily generate many-body
interactions that significantly complicate the calculations and are
often truncated for that reason.  Renormalization without retaining
the effective many-body potentials 
abandons the variational upper bound characteristic that we prefer to
retain.  Furthermore, convergence with increasing model space is
generally neither uniform nor monotonic when applying renormalization
without retaining the induced many-body potentials.  This leads to
challenges for extrapolation to infinite model spaces.

Our NCFC approach requires methods to
reliably extrapolate results obtained in a finite basis space to the
infinite or complete basis space limit.  This need for extrapolation
tools mirrors similar situations in other fields of science where a
sequence of results with increasing resolution must be extrapolated to
the limit of infinite resolution.
The resulting high precision results of the NCFC do not agree exactly
with experiment. Indeed, no known realistic NN interactions provide exact
descriptions of a similar range of nuclear data that we examine and
it is probable that NNN and higher body interactions are
needed.

\begin{figure}
\centerline 
{\includegraphics[width=9cm]{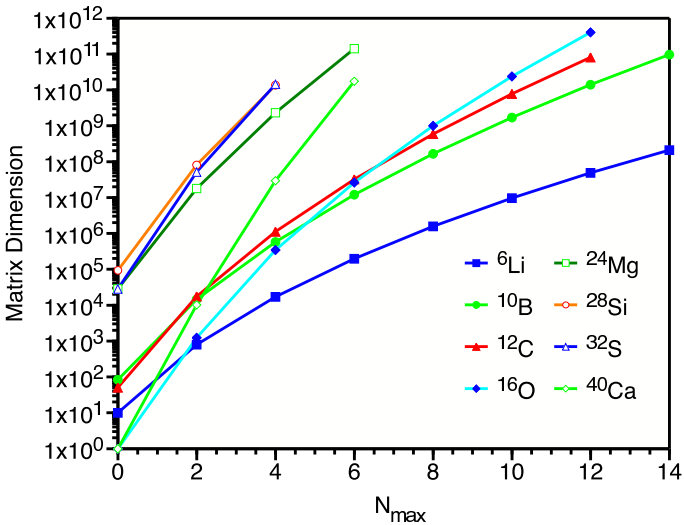}}
\caption{(Color online) Representative Hamiltonian matrix dimensions
for total magnetic projection $M =0$ states in the single-particle
$m$-scheme as a function of the maximum total oscillator quanta 
of excitations
specified by $N_{max}$.  The natural parity matrix dimensions are
represented by the specific points while unnatural parity matrix
dimensions would lie close to the interpolating lines at odd values of
$N_{max}$.}
\label{Dimensions}
\end{figure}

In order to further motivate our efforts to develop robust
extrapolation tools, we show in Fig.~\ref{Dimensions} the Hamiltonian
matrix dimensions for a set of representative light nuclei.  
We employ
the ``$m$-scheme" where each HO single-particle state has its orbital and spin
angular momenta coupled to good total angular momentum, $j$, and magnetic
projection, $m$.  
The many-body basis states are Slater determinants in this HO basis and
are limited by the imposed symmetries --- parity
and total angular momentum projection ($M$), as well as by $N_{max}$.  
In the natural parity cases for even nuclei shown, $M=0$
enables the simultaneous calculation of the entire spectrum for that
parity and $N_{max}$.

The nearly exponential growth in matrix dimension with increasing
$N_{max}$ is clearly evident in Fig.~\ref{Dimensions}.  In order to
achieve NCFC results for the heavier of these nuclei by extrapolation, using
realistic interactions, we would need to diagonalize matrices that are
beyond the reach of present technologies.  However, in cases up to and
including $^{16}$O, we may expect to obtain systematic results for the
first few increments of $N_{max}$.  In order to utilize the sequence of results obtained 
with $N_{max}$ values that are currently accessible, we need to investigate
suitable extrapolation tools.

\begin{figure}
\centerline 
{\includegraphics[width=9cm]{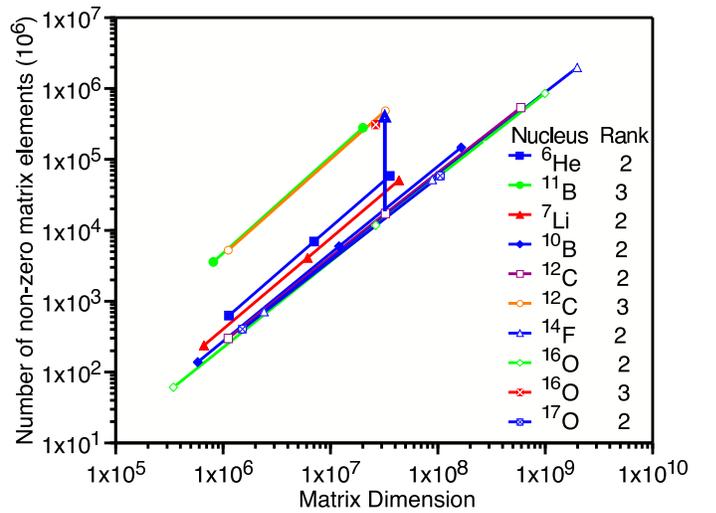}}
\caption{(Color online) Number of non-vanishing many-body Hamiltonian
matrix elements for representative light nuclei as a function of the
basis space dimension.  The points represent sample cases that have
been solved and correspond to those indicated in the legend.  The
curves approximate a $D^{3/2}$ power law where $D$ is the basis space
dimension. The vertical arrow measures a factor of 30 between the
2-body Hamiltonian (rank=2) and 3-body Hamiltonian (rank=3) cases for
$^{12}$C at the same dimension corresponding to $N_{max}=6$.  Note the
logarithmic scales.}
\label{Nonvanish}
\end{figure}

To better understand the scale of computational effort needed for
no-core microscopic nuclear structure studies, 
we consider the
memory storage demands as a function of matrix dimension.
For several representative nuclei, 
we enumerate the number of
non-vanishing matrix elements of the resulting many-body 
Hamiltonian matrix (its lower triangle
only for efficiency) and display
the resulting counts as a function of the matrix dimension in
Fig.~\ref{Nonvanish}.  We present results for the case of a 2-body
input Hamiltonian (NN interaction only) and for the case of a 3-body
Hamiltonian (NN + NNN interactions).  In spite of the very large
memory requirements, the various curves display an encouraging trend.
Specifically, the number of non-vanishing many-body matrix elements follows a
$D^{3/2}$ growth rate, where $D$ is the dimension of the matrix.  That
is, the matrices exhibit a very sparse character and this is the
property that allows us to diagonalize the large matrices that we can
presently solve.

\section{Selection of Hamiltonian ingredients and example nuclei}

In order to carry out the NCFC calculations, we require a
realistic NN interaction that is sufficiently weak at high momentum
transfers that we can obtain a reasonable convergence trend.  The
conventional Lee--Suzuki--Okamoto renormalization procedure of the {\it
ab initio} NCSM \cite{NCSMC12} develops soft, $N_{max}$-dependent, 
effective interactions that provide answers close to experimental observations.  
However, the convergence trend of the results with
increasing $N_{max}$ is often not uniform and leads to challenges for
extrapolation to infinite model spaces.  Nevertheless, there is also 
encouraging progress in extrapolating NCSM ground state energies
of light nuclei using different strategies \cite{Zhan03,Forssen08}.  
Of course, as the basis space increases, one expects the NCSM and NCFC methods to
arrive at the same exact result.  Thus, the choice of method, NCSM or NCFC, 
will ultimately depend on the 
underlying Hamiltonian selected for the application.
In the NCFC approach discussed here, we seek to obtain the ground
state energy of the original, or ``bare" \cite{bare},
Hamiltonian in the infinite model
space with evaluated uncertainties.  To this end, 
we incorporate systematic and reliable extrapolation tools as needed.

\begin{figure}
\centerline 
{\includegraphics[width=9cm]{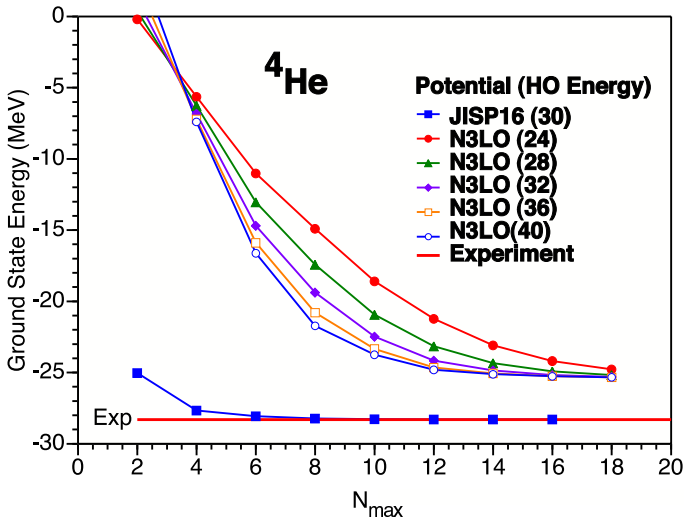}}
\caption{(Color online) Calculated ground state energy of $^4$He 
as a function of $N_{max}$ at various values of
the oscillator energy, $\hbar\Omega$, as indicated in the legend.  The
results are connected by straight line segments to guide the eye.  The
results with chiral N3LO are from Ref.~\cite{Navratilpc}.
The results for JISP16 are closer to convergence even in rather modest
basis spaces. 
No extrapolations are needed in these cases as converged results are obtained directly.}
\label{4He_JISP16_v_N3LO}
\end{figure}

We compare in Fig.~\ref{4He_JISP16_v_N3LO} the ground state of $^4$He
using the chiral interaction, N3LO
\cite{N3LO,Navratilpc}, with JISP16 \cite{Shirokov07,JISP16_web} and plotted as a
function of $N_{max}$ to ascertain convergence rates.  
All our results include the Coulomb interaction between protons.
Each point represents the ground state energy from an 
$N_{max}$-truncation of the full infinite matrix problem.  Hence,
all points are
strict upper bounds on the exact ground state energies for the
respective Hamiltonian.  Fig.~\ref{4He_JISP16_v_N3LO} shows that
JISP16 provides a faster convergence rate and a ground state energy in
closer agreement with the experimental energy of $-28.296~\rm{MeV}$.
JISP16 also produces spectra and other observables in light nuclei
that are in reasonable accord with experiment \cite{Shirokov07}.
Indeed, this interaction was designed to possess these specific
properties while retaining an excellent description of the NN data.
The interaction is non-local but this is no limitation for NCFC
which also preserves all the symmetries of the underlying NN interaction.

With JISP16 for our NN interaction, we perform {\it ab initio} calculations of
the ground state energies of $^2$H, $^3$H, $^4$He, $^6$He, $^6$Li,
$^8$He, $^{12}$C and $^{16}$O.  The three lightest nuclei serve as
test cases to demonstrate that the extrapolation methods, using
results in limited basis spaces, are able to predict the fully
converged results and to demonstrate that our assessed uncertainties
are realistic.  We limit ourselves to examples for which a sufficient
set of results could be achieved within our current computational
resource limits. 

\section{Finite basis space expansions}

Our results in finite basis spaces satisfy the variational principle and
show uniform and monotonic convergence from above to the exact eigenenergy
with increasing $N_{max}$.  That is, the results for the energy of the
lowest state of each spin and parity, at any $N_{max}$ truncation, are
upper bounds on the exact converged answers and the convergence is
monotonic with increasing $N_{max}$.  This guarantee of monotonic
convergence from above to the exact energy facilitates our choice of
extrapolating function. 

We carefully investigate the dependence of the results on the basis
space parameters, $N_{max}$ and $\hbar\Omega$.  Our goal is to achieve
independence of both of these parameters as that is a signal for
convergence --- the result that would be obtained from solving the same
problem in a complete basis.

Before proceeding, let us explain some additional features of the many-body regulator, 
$N_{max}$.  
As introduced above, 
$N_{max}$ is the maximum number of oscillator quanta
shared by all nucleons above the lowest HO configuration for the
chosen nucleus.  Its use allows us to factorize eigenfunctions into
intrinsic and center-of-mass (CM) components for ease of eliminating
spurious CM motion effects on all observables.
One unit of oscillator quanta is one unit of the quantity $(2n+l)$
where $n$ is the principle quantum number and $l$ is the 
angular quantum number.
If the highest HO single-particle state of this lowest HO
configuration has $N_0$ HO quanta, then $N_{max}+N_0$ identifies the
highest HO single-particle states that can be occupied within this
many-body basis.
Note that since
$N_{max}$ is the maximum of the {\em total} HO quanta above the
minimal HO configuration, we can have at most one nucleon in such a
highest HO single-particle state.

The precise method of achieving the factorization of the CM and intrinsic
components of the many-body wavefunction follows a standard approach, sometimes
referred to as the ``Lawson'' method \cite{Lawson}.  
In this method, one selects the many-body basis space in the manner described above and adds
a Lagrange multiplier term to the many-body Hamiltonian 
$\lambda(H_{CM} - \frac{3}{2} \hbar\Omega)$
where $H_{CM}$ is the Harmonic Oscillator Hamiltonian for the CM motion.  With $\lambda$
chosen positive (10 is a typical value), one separates the states of lowest CM motion 
($0S_{\frac{1}{2}}$)
from the states with excited CM motion by a scale of order $\lambda \hbar\Omega$.  The 
resulting low-lying states have wavefunctions that are assured to have the desired factorized form.

It is important to note that our NCFC results for the ground
state energy for $A=2, 3~\&~4$ are obtained directly as we achieve sufficient
independence of $N_{max}$ and $\hbar\Omega$.  For the other nuclei
studied here, we characterize the approach to convergence by the
dependence of results on both $N_{max}$ and $\hbar\Omega$ and
investigate the shape of that convergence in detail.  
The degree of
residual dependence on these two parameters provides a measure of the
difference from the exact result, an estimate of the numerical
uncertainty in the extrapolation.

We employ the parallel-processor code 
``Many-Fermion Dynamics --- nuclear" (MFDn)
\cite{Vary92_MFDn} that sets up the many-body basis space,
evaluates the many-body Hamiltonian matrix, 
obtains the low-lying eigenvalues and eigenvectors using the
Lanczos algorithm, and evaluates a suite of experimental observables.  
Working in the single-particle HO $m$-scheme,
the lowest 15 states here are usually obtained with $300{-}600$
iterations, depending on $N_{max}$ and the nucleus involved.  The
required number of iterations grows with $N_{max}$.

The largest matrix we diagonalize for this work corresponds to $^{16}$O in the
$N_{max}=8$ space with a basis dimension about 1 billion.  We obtain the
lowest 8 eigenstates and a suite of observables in 4.5 hours on 12,090
processors using the Franklin supercomputer at the National Energy
Research Supercomputer Center (NERSC).  The second largest case is
$^{12}$C with a basis dimension of about 600 million for which we obtain the
lowest 15 eigenstates and a suite of observables in 2.3 hours on
12,720 processors using the Jaguar supercomputer at Oak Ridge National
Laboratory (ORNL).  The above times correspond to calculations at a
single value of $\hbar\Omega$.  For calculations as a function of
$\hbar\Omega$ in the same basis spaces, we use internally generated
and stored index arrays amounting to many terabytes of data so that
the second and subsequent $\hbar\Omega$ values each take about 2/3 the
time as the first case.

Following the completion of the calculations reported here, further
speedups have been accomplished with the code so that the
above-mentioned times are reduced by a factor of 2 in future
calculations of the same type \cite {Sternberg08}.

\section{Extrapolating the ground state energy}
\subsection{Simple illustration}

Here we illustrate the convergence properties for the nuclear ground
state energy in a HO basis with a simple model.  While the properties
of the HO basis are useful for many purposes, such as the exact
treatment of the center of mass motion and the ease of transforming
between relative and single-particle coordinates, the asymptotic HO
wavefunctions are gaussians while wavefunctions of finite nuclei will
have exponentially decreasing amplitudes at large distances.  Correct
long-range behavior is important for precision evaluation of energies
and for many other experimental properties such as electromagnetic
moments and transition rates.  To achieve converged long-range
observables, we expect to require an optimal choice of the
$\hbar\Omega$ value and sufficiently large values of $N_{max}$ to
generate good asymptotic properties.  To investigate these issues, we
evaluate the properties of fermions in a finite phenomenological
potential, the picture that underlies the successful nuclear shell
model, using a HO basis expansion.

Consider the properties of a single Slater determinant composed of the
lowest $A$-particle orbits of the Saxon--Woods central potential plus a
nuclear spin-orbit potential.  This corresponds to a standard mean
field description of the nucleus with $A$ nucleons and we refer to
this simple model as the ``Extreme Single Particle Model" (ESPM).
Instead of solving for the single-particle states by numerical
integration, we diagonalize the 1-body model Hamiltonian in a HO basis
to simulate the procedures of a no-core finite HO basis calculation.  

We adopt
$^{12}$C as an example and we perform this diagonalization as a
function of $N_{max}$ and $\hbar\Omega$.  In this way, we are studying
how the lowest $s$-state and $p$-state solutions depend on $N_{max}$ and
$\hbar\Omega$.  We identify $N_{max}$ with the maximum value of the HO
quanta $(2n+l)$ retained in the HO basis expansion so that $N_{max}=0$
corresponds to a pure HO approximation, $N_{max}=2$ employs 2 basis HO
functions for the occupied 0s-states and 0p-states, $N_{max}=4$
employs 3 basis HO functions for the occupied single-particle states,
etc.  Note, there is a difference between the use of $N_{max}$ in this
model problem from our NCFC approach.  Here, since we work
entirely in a single-particle basis, all particles have
simultaneous access to the range of basis states dictated by $N_{max}$
while in our NCFC, the many-body basis is restricted so that as
one particle takes more quanta, the remaining particles take fewer
quanta.  

It should be noted that our model problem more closely simulates the
traditional ``Full Configuration Interaction" (FCI)  approach used in quantum chemistry where orbits are
equally accessible in the many-body basis states up to some
single-particle cutoff.  The origin of this difference is our need to
retain an exact treatment of the center-of-mass motion in our no-core methods
for finite nuclei (NCFC and NCSM).  In spite of this difference in cutoffs, we find
that this simple model is useful for illustrating how proper asymptotic wave function
properties influence convergence rates for self-bound nuclei in the HO basis.

We adopt a Saxon--Woods central, $U(r)$, and spin-orbit, $U_{so}(r)$,
potentials similar to a standard choice \cite{Perey} where, for
simplicity, we use the same parameters for the neutrons and the
protons of $^{12}$C
\begin{eqnarray*}
U(r) &=&  \frac{U_0}{1+\exp((r-R)/a_0)} \\
U_{so}(r) &=& \vec{S} \cdot
          \vec{L} \; \left( \frac{\hbar}{m_{\pi}c^2}\right)^2 
             \frac{1}{r} \frac{d}{dr} \frac{U_{SO}}{1+\exp((r-R)/a_{SO})}
\end{eqnarray*}
with $R = r_0  A^{1/3} $ and 
$\left(\frac{\hbar}{m_{\pi}c^2}\right)^2 = 2.0 ~{\rm fm}^2$.
Our parameters selected for this demonstration are $U_0~=~-32$~MeV,
$r_0~=~1.25$~fm, $a_0~=~0.65$~fm, $U_{SO}~=~15$~MeV and
$a_{SO}~=~0.47$~fm.  For the protons we add the Coulomb field of a
uniformly charged sphere of radius $R$.

\begin{figure}
\centerline 
{\includegraphics[width=9cm]{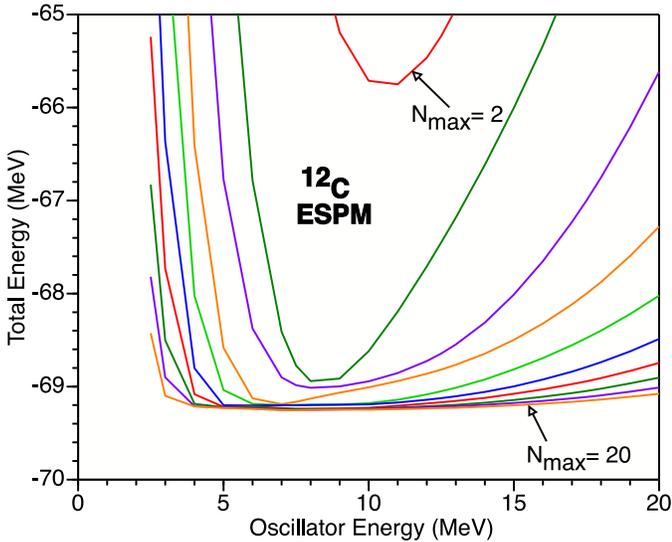}}
\caption{(Color online) Calculated ground state energy of $^{12}$C in
the Extreme Single Particle Model (ESPM) as a function of
$\hbar\Omega$ and $N_{max}$, the maximum value of the HO states'
quanta, $2 n + l$, used in expanding the Saxon--Woods single-particle
states.  The curve closest to convergence corresponds to the value
$N_{max} = 20$ and successively higher curves are obtained with
$N_{max}$ decreased by 2 units for each curve.}
\label{12C_Saxon_Woods_gs_energy}
\end{figure}

We add the energies of the occupied orbits, taking into account
degeneracies, to obtain the total energy of the system in the ESPM.
The resulting ESPM ground state energy for $^{12}$C is displayed in
Fig.~\ref{12C_Saxon_Woods_gs_energy} as a function of $\hbar\Omega$
for a range of $N_{max}$ values up to and including $N_{max} = 20$.
The line segments connect the results calculated at selected values of
$\hbar\Omega$.  Here we observe a pattern that is typical of our
no-core basis space results presented below --- a sequence of curves with energy
decreasing as a function of increasing $N_{max}$, consistent with the
property dictated by the variational principle.  With increasing
$N_{max}$, one approaches convergence signaled by achieving
simultaneous independence of both $N_{max}$ and $\hbar\Omega$.  In the
ESPM, we achieve the total energy converged to within 10 keV at
$N_{max}=20$ over the range $\hbar\Omega = 6{-}11$ MeV.
For the optimal value of $\hbar\Omega = 7$ MeV, we achieve a total
energy to within 170 keV of the exact answer already at $N_{max} = 8$,
yielding an upper bound within 0.25\% of the exact result.

\begin{figure}
\centerline
{\includegraphics[width=9cm]{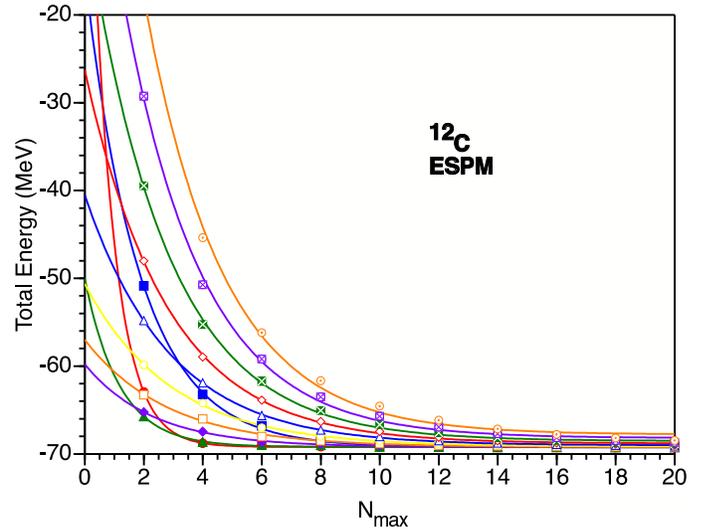}}
\caption{(Color online) Calculated ground state energy of $^{12}$C in
the Extreme Single Particle Model (ESPM) as a function of $N_{max}$
for selected values of $\hbar\Omega$ used in defining the basis
states.  The points correspond to $\hbar\Omega$ values ranging from 5
to 30 MeV in 2.5 MeV increments.  The curves represent an exponential
plus a constant fit to each set of results at fixed $\hbar\Omega$,
excluding the $N_{max} = 0$ result.  Each point carries equal weight.}
\label{12C_Saxon_Woods_gs_extrap}
\end{figure}

For the purpose of exploring potential extrapolation tools, we use the
results of the ESPM to map out the convergence pattern of the total
energy in the present work.  In subsequent efforts, we will
investigate other observables in a similar fashion.  Thus, we present
in Fig.~\ref{12C_Saxon_Woods_gs_extrap} the total energy as a function
of $N_{max}$ at fixed values of $\hbar\Omega$ spanning the minima of
the curves in Fig.~\ref{12C_Saxon_Woods_gs_energy}.  Specifically, the
points correspond to $\hbar\Omega$ values ranging from 5 to 30 MeV in
2.5 MeV increments.  We find that, once we exclude the $N_{max} = 0$
result, the calculated points appear to represent an exponential
convergence pattern.  To confirm this, we fit an exponential plus
constant to each set of results as a function of $N_{max}$, excluding
$N_{max} = 0$, and the resulting fits are displayed in
Fig.~\ref{12C_Saxon_Woods_gs_extrap} as smooth curves.  That is, for each set of points
at fixed $\hbar\Omega$, we fit the ground state energy with three
adjustable parameters using the relation
\begin{equation}
E_{gs}(N_{max}) = a \exp(-c\,N_{max}) + E_{gs}({\infty}).
\label{extreq}
\end{equation}
In these fits, we assigned equal weight to each point and perform a
regression analysis.

Overall, we conclude that the exponential plus constant fits the 
results rather well.  Thus, one observes that the HO basis provides a
rapidly converging sequence of total energies in the ESPM, one
well-represented by exponential convergence in $N_{max}$ towards the
the asymptotic total energy, $E_{gs}({\infty})$.  It appears reasonable to expect 
this convergence pattern of the HO basis treatment of the ESPM to be representative 
of HO basis expansion behavior in our no-core applications, we will
adopt this functional form as a foundation for further developing our
extrapolation methods below.  In the following sections, we will use
additional arguments to improve upon this tool and test it in light
nuclei where converged results are obtained directly.

We note that a similar exponential behavior for HO basis space calculations of
a cold trapped Fermi gas has been observed \cite {Alhassid08}.  In that case, the
same type of single-particle-space regulator was employed as we use here
in the ESPM application.

The exponential plus constant has also been employed as an extrapolation
tool in more conventional shell model studies as well \cite{horoi}.
In those applications, the variable is the matrix dimension rather than
$N_{max}$.

\subsection{NCFC test case: deuteron}

Next, we turn to the NCFC calculations for light nuclei using
JISP16, where we can achieve nearly exact results in large model
spaces.  In this and the following subsections, we investigate the convergence 
rates for the ground state energies as a function 
of $N_{max}$ and $\hbar\Omega$ for $^2$H, $^3$H,
and $^4$He.  We discuss two extrapolation methods, which allow us to
obtain estimates of the converged NCFC results from finite model
spaces. We also introduce the assessed uncertainties for our extrapolated results.

\begin{figure}
\centerline 
{\includegraphics[width=9cm]{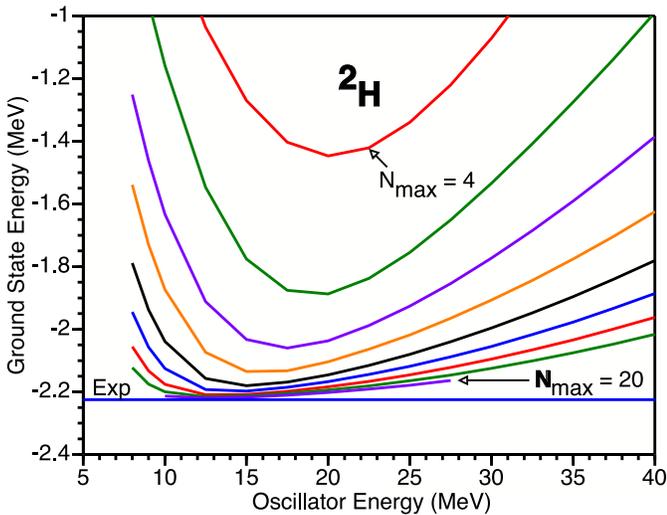}}
\caption{(Color online) Calculated ground state energy of $^2$H as a
function of the oscillator energy, $\hbar\Omega$, for selected values
of $N_{max}$ used in defining the basis states.  The curve closest to
experiment corresponds to the value $N_{max} = 20$ and successively
higher curves are obtained with $N_{max}$ decreased by 2 units for
each curve. The curves are formed by straight line segments joining calculated results.}
\label{2H_gs_vs_hw}
\end{figure}

\begin{figure}
\centerline 
{\includegraphics[width=9cm]{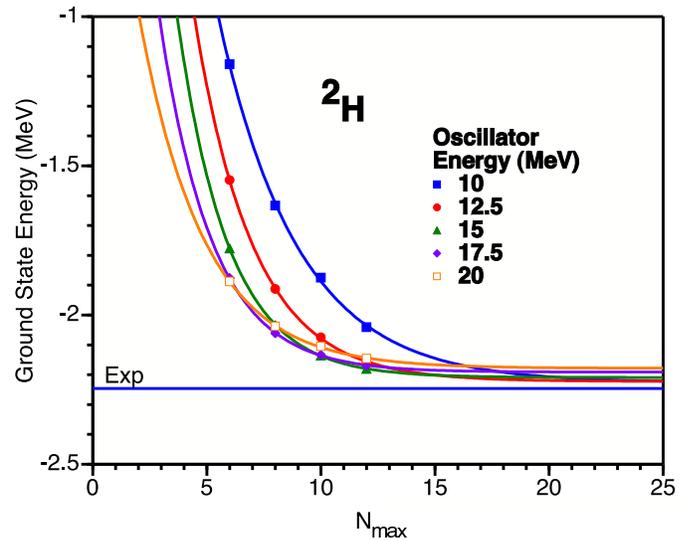}}
\caption{(Color online) Calculated ground state energy of $^2$H as a
function of $N_{max}$ at $\hbar\Omega$ values that bracket the best
upper bound as described in the text.  The smooth curves are fits by
Eq. (\ref{extreq}) 
to the 4 data points shown at each
value of $\hbar\Omega$ with each point weighted equally.}
\label{2H_gs_extrap}
\end{figure}

The sequence of curves in Fig.~\ref{2H_gs_vs_hw} for $^2$H illustrates
the trends we encounter in our calculations when evaluating
the ground state energy with JISP16.  
Our purpose with $^2$H is only to illustrate convergence trends and
test the extrapolation tool since the exact answer is also available
from the direct solution of the Schroedinger equation
\cite{Shirokov05}, and agrees with experiment.  The $N_{max} = 18$
curve reaches to within 9 keV of this exact result; the $N_{max} = 20$
curve reaches to within 5 keV.  We note that the weak binding of $^2$H
leads to a slow progression of the curves towards independence of
$\hbar\Omega$ and contrasts the stronger binding situation obtained
for $^4$He shown below in Fig.~\ref{4He_gs_vs_hw}.

\subsubsection{Global extrapolation method}

We use these $^2$H results to test our ``global extrapolation method"
(referred to as ``extrapolation A") as illustrated in
Fig.~\ref{2H_gs_extrap}.  Here, we fit only four calculated points at
each value of $\hbar\Omega$ in the range $N_{max} = 6 {-}12$,
representative of the limited results that we expect to encounter in
slightly heavier systems.  We select the values of $\hbar\Omega$ to
include in the analysis by first taking the value at which the minimum
(with respect to $\hbar\Omega$) occurs along the highest $N_{max}$
curve included in the fit, then taking one $\hbar\Omega$ value lower
by 2.5 MeV and three $\hbar\Omega$ values higher by successive
increments of 2.5 MeV.  Since the minimum occurs along the
$N_{max}=12$ curve at $\hbar\Omega = 12.5$ MeV as shown in
Fig.~\ref{2H_gs_vs_hw}, this produces the 5 curves spanning a range of
10 MeV in $\hbar\Omega$ shown in Fig.~\ref{2H_gs_extrap}.
We perform a linear regression for each sequence of four points at
fixed $\hbar\Omega$ using an independent exponential plus constant for
each sequence, and observe a small spread in the extrapolants as is
evident in Fig.~\ref{2H_gs_extrap}, which is indicative of the
uncertainty in this method.

We recognize that this window of results in $\hbar\Omega$ values is
arbitrary.  Our only assurance is that it seems to provide a
consistent set of extrapolations in the nuclei examined in the present
work.

For the global extrapolation we chose sets of four points due to a
desire to minimize the fluctuations arising from certain ``odd-even" effects
already visible in Fig.~\ref{12C_Saxon_Woods_gs_energy}.  These
effects are most pronounced in weakly bound systems, and may be
attributed to the fact that HO wavefunctions fall off too fast:
wavefunctions of finite nuclei decrease exponentially at large
distances.  In order to mimic such an exponential decrease with HO
basis functions, one needs HO basis functions with both even principle
quantum number $n$ (even number or nodes in radial wavefunction) and
with odd principle quantum number $n$ (odd number or nodes in radial
wavefunction).  Since $N = (2n+l)$, a set of four successive $N_{max}$
points (with $N_{max}$ even) implies 
we incorporate two highest allowed HO single-particle states 
with even values of the principal quantum number and two highest
allowed with odd values.
Thus, a set of four
consecutive $N_{max}$ points instead of three points (the minimal
number of points for an exponential extrapolation) averages out some
of these ``odd-even'' effects.  We will come back to this point when
we discuss the extrapolation method B, using sets of three points at
fixed $\hbar\Omega$.

Next, we consider what weight to assign to each calculated
$(N_{max},\hbar\Omega)$-point.
The fits in Fig.~\ref{2H_gs_extrap} are obtained with equal weights
for each of the points.  However, as $N_{max}$ increases, we are
approaching the exact result from above with increasing precision.
Hence, the importance of results grows with increasing $N_{max}$ and
this should be reflected in the weights assigned to the calculated
points used in the fitting procedure.  With this in mind, we adopt the
following strategy: define a chisquare
function to be minimized and assign a $\sigma_{N_{max}}$ to the gs energy at each
$N_{\max}$ value that is the change in the calculated
energy from the next lower $N_{\max}$ value 
 
$$\sigma_{N_{max}}=(E_{gs}(N_{max})-E_{gs}(N_{max}-2))$$.

To complete these sigma
assignments, the sigma for the lowest $N_{\max}$ point on the 
$N_{\max}$ curve is assigned a value three times the sigma 
calculated for the second point on the same fixed-$\hbar\Omega$ trajectory.  
As a final element to our global extrapolation strategy, we invoke the
minimization principle to argue that all curves of results at fixed
$\hbar\Omega$ will approach the same exact answer from above.  Thus
all curves will have a common asymptote and we use that condition as a
constraint on the chisquare minimization.

When we use exponential fits constrained to have a common asymptote
and weights based on the local slope, we obtain curves close to those
in Fig.~\ref{2H_gs_extrap}.  The differences are difficult to perceive
in a graph so we omit presenting a separate figure for them in this
case. It is noteworthy that the equal weighting of the linear
regression leads to a spread in the extrapolants that is still modest.

\begin{figure}
\includegraphics[width=0.99\columnwidth]{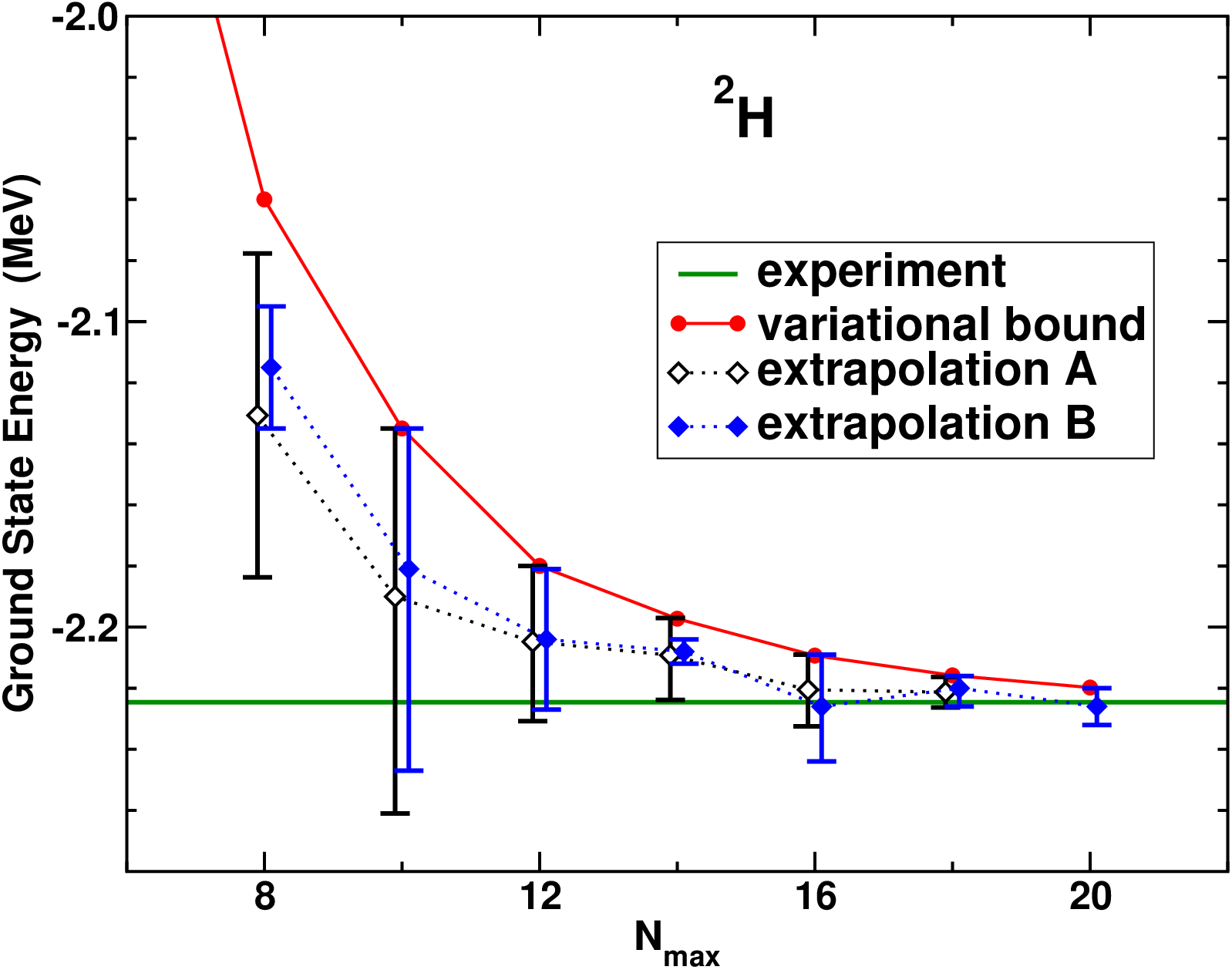}
\caption{(Color online) Extrapolated ground state energies and
variational upper bounds from each set of four (extrapolation A) or
three (extrapolation B) successive $N_{max}$ values as a function of
the largest value of $N_{max}$ in each set.  Error bars are dominated
by the uncertainties in the extrapolations, and are obtained as described
in the text.  Note the expanded scale and the reasonable consistency
of the extrapolated results: for $N_{max} \ge 10$ all but one are
within their uncertainty range of the exact answer.
\label{2H_gs_extrap_insert}}
\end{figure}

The sequence of asymptotes for the $^2$H ground state energy, obtained
with the global extrapolation, by using successive sets of four points
in $N_{max}$ and performing our constrained fits to each such set of
four points, is shown in Fig.~\ref{2H_gs_extrap_insert} as extrapolation A.  
We employ
the independent fits such as those in Fig.~\ref{2H_gs_extrap} to
define the uncertainty in our asymptotes.  In particular, we define
our uncertainty, or estimate of the standard deviation for the
constrained asymptote, as one-half the total spread in the asymptotes
arising from the independent fits with equal weights for each point.
On rare occasions, we obtain an outlier when the linear regression
produces a residual less than 0.999 that we discard from the
determination of the total spread.  Also, on rare occasions, the
calculated upper uncertainty reaches above the calculated upper bound.
When this happens, we reduce the upper uncertainty to the upper bound
as it is a strict limit.

One may worry that the resulting extrapolation tool contains several
arbitrary aspects and we agree with that concern.  One recourse
is to cross-check these choices with the solvable NCFC cases in
the present subsection and following subsections.  
We seek consistency of the constrained
extrapolations as gauged by the uncertainties estimated from the
unconstrained extrapolations described above.  Indeed, our results
such as those shown in Fig.~\ref{2H_gs_extrap_insert}, demonstrate
that consistency.  The deviation of any specific constrained
extrapolant from the result at the highest upper limit $N_{max}$
appears well characterized by the assigned uncertainty.

\subsubsection{Extrapolation at fixed $\hbar\Omega$}

In addition, we also employ an extrapolation at fixed values of
$\hbar\Omega$ using only three successive values of $N_{max}$, the
minimal number of points for such an extrapolation (referred to as
extrapolation B).  Under the assumption that the convergence is indeed
exponential, such an extrapolation should get more accurate as
$N_{max}$ increases; the difference between the extrapolated results
from two consecutive sets of three $N_{max}$ values is used here as our
estimate of the numerical uncertainty associated with the
extrapolation.  

\begin{figure}
\includegraphics[width=0.99\columnwidth]{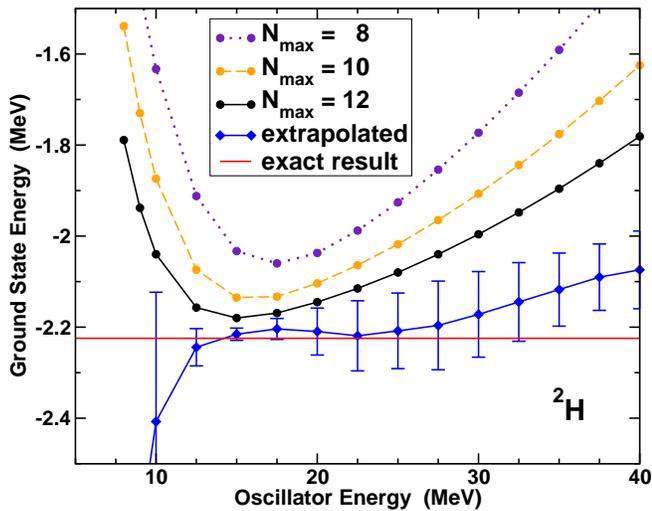}
\caption{(Color online) Calculated ground state energy of $^2$H for
$N_{max} = 8$, $10$, $12$, and the extrapolated ground state energy
using method B, as function of the oscillator energy, $\hbar\Omega$.
Error bars are obtained from the difference with the extrapolation
using $N_{max} = 6$, $8$, and $10$ calculations.
\label{2H_gs_extrapBtest}}
\end{figure}

In Fig.~\ref{2H_gs_extrapBtest} we illustrate this extrapolation for
$^2$H based on calculations with $N_{max}=8$, $10$, and $12$.  As
we can see, this extrapolation gives $\hbar\Omega$ dependent results.
We therefore consider the value of $\hbar\Omega$ where the extrapolation is most stable (i.e. for which the difference between the extrapolated value and the result at the highest $N_max$ is minimal) 
as the best or most reliable  $\hbar\Omega$ for this extrapolation method.
This $\hbar\Omega$ value is usually at or slightly above the variational minimum.

Since this extrapolation uses sets with only three $N_{max}$ points,
the ``odd-even'' effects may be significant, in particular for weakly
bound nuclei.  This is indeed what we find for $^2$H as seen in
Fig.~\ref{2H_gs_extrap_insert}.  Nevertheless, within the estimated
error bars, the results are consistent with extrapolation method A and
with the exact result.  In addition, as we proceed to applications in
heavier nuclei and more deeply bound nuclei, this extrapolation
becomes more stable and useful, as we will see below.

\subsection{\boldmath More NCFC test cases: $A = 3,\,4$}

\begin{figure}
\centerline 
{\includegraphics[width=9cm]{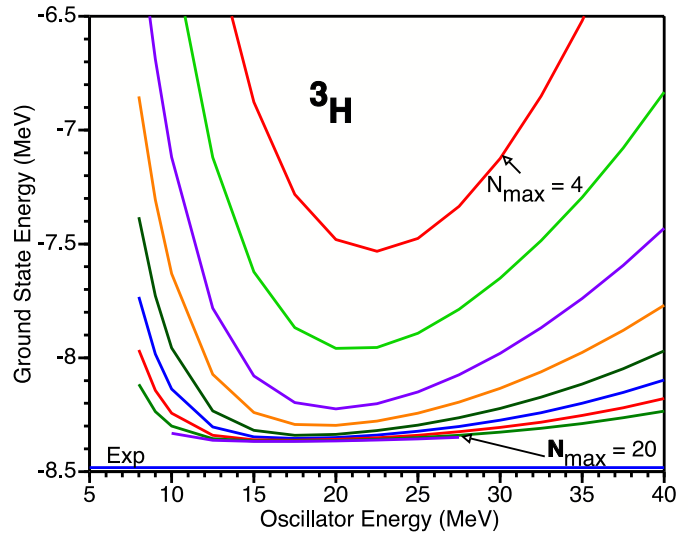}}
\caption{(Color online) Calculated ground state energy of $^3$H as a
function of the oscillator energy, $\hbar\Omega$, for selected values
of $N_{max}$.  The curve closest to experiment corresponds to the
value $N_{max} = 20$ and successively higher curves are obtained with
$N_{max}$ decreased by 2 units for each curve. }
\label{3H_gs_vs_hw}
\end{figure}

\begin{figure}
\centerline 
{\includegraphics[width=9cm]{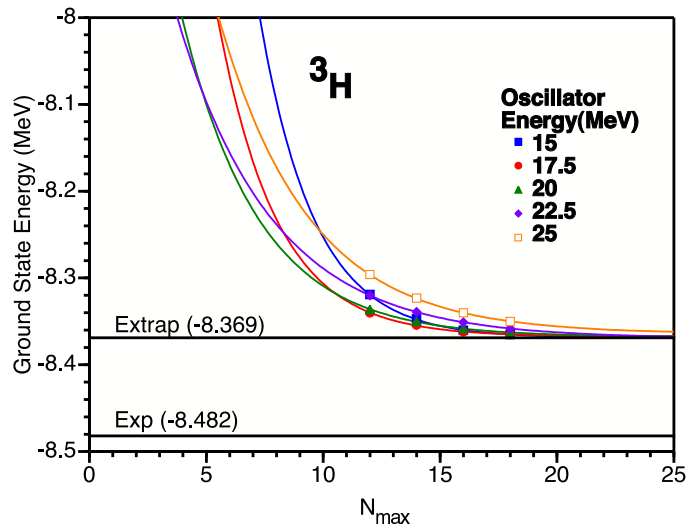}}
\caption{(Color online) Calculated ground state energy of $^3$H as a
function of $N_{max}$ at $\hbar\Omega$ values that bracket the best
upper bound. 
The smooth curves are fits of an exponential plus a constant [see
  Eq. (\ref{extreq})] 
to the 4
data points shown at each value of $\hbar\Omega$.  There is a 5 keV
spread in these asymptotes that is used to assign the uncertainty to
the asymptote derived from the constrained fit as described in the
text.  The asymptote of the constrained global fit (extrapolation A)
is quoted in the figure and the experimental result is shown for
comparison.}
\label{3H_gs_extrap}
\end{figure}

\begin{figure}
\includegraphics[width=0.99\columnwidth]{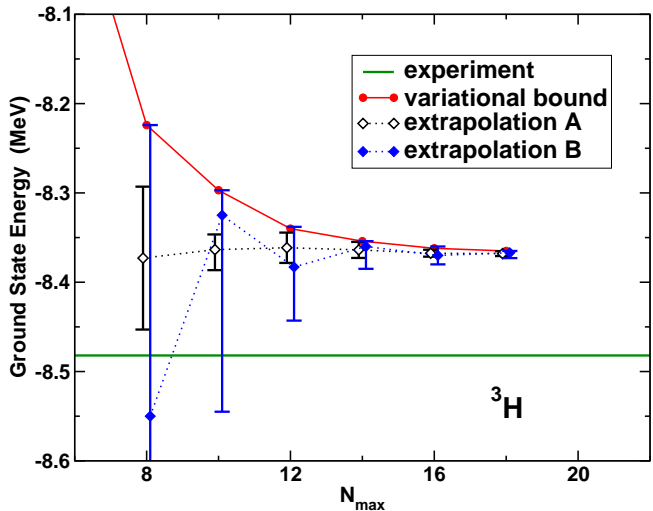}
\caption{(Color online) Extrapolated ground state energies and
variational upper bounds from each set of four (extrapolation A) or
three (extrapolation B) successive $N_{max}$ values as a function of
the largest value of $N_{max}$ in each set.  Error bars are dominated
by the uncertainties in the extrapolations, and obtained as described
in the text.  Note the consistency of the extrapolated results.
\label{3H_gs_extrap_insert}}
\end{figure}

The ground state energies of $^3$H using JISP16 are shown in
Fig.~\ref{3H_gs_vs_hw} as a function of the HO energy for the same
sequence of basis spaces as for $^2$H.  We again observe a converging
sequence of upper bounds with an indication of a small amount of
underbinding compared with experiment.  We note that the curves show a
greater region of approximate independence of $\hbar\Omega$ than found
in the case of $^2$H as may be expected from the stronger binding
relative to the first breakup threshold in the present situation.

We use the case of $^3$H to illustrate again the utility of the global
extrapolation (A).  Results of independent fits with equal weights for
each calculated point are shown in Fig.~\ref{3H_gs_extrap} to
demonstrate the nearly identical asymptote when results are available
at sufficiently high $N_{max}$ values.  We depict in
Fig.~\ref{3H_gs_extrap} both the experimental value and the asymptote
of the global extrapolation using the four point sequence up to 
$N_{max}=18$.  
The span of $\hbar\Omega$ values is selected
in the same manner as in the $^2$H case, the procedure we will use
throughout this work.  

As for the case of $^2$H, we also present the sequence of extrapolated
results for the $^3$H ground state energy in
Fig.~\ref{3H_gs_extrap_insert} using both extrapolation methods A and
B, together with the variational bound.  Both extrapolation methods
appear to be consistent with each other, and give numerical error bars
which decrease with increasing $N_{max}$.  The extrapolation B, using
only three successive $N_{max}$ points at fixed $\hbar\Omega$, shows a
rather strong ``odd-even'' effect.  Nevertheless, all extrapolated
results agree, within error bars, with each other, and with our best
results at $N_{max}=20$.  These results are also quoted in
Table~\ref{tabA_2-16}.  We conclude that JISP16 underbinds $^3$H by
approximately 113~keV.

Our calculations for $^3$He show a similar convergence
pattern as those for $^3$H.  At $N_{max}=20$, our results are
within a few keV of full convergence, as can be seen from
Table~\ref{tabA_2-16}, and we find JISP16
underbinds $^3$He by about
52~keV.

\begin{table*}
\renewcommand{\arraystretch}{1.3} 
\caption{Binding energies in MeV
of nine nuclei and of the first excited $0^+$ states in $^{12}$C and
$^{16}$O from experiment and theory.  The experimental values are from
Refs.~\protect\cite{AS88,Fujita04,AS90,T88,OZ88,AS91}.  
The uncertainties in the rightmost digits of an extrapolation are
quoted in parenthesis.  The bounds for the binding energies
follow from the variational upper bounds for the ground state
energies.
 The rightmost column provides the uppermost value of
$N_{\max}$ used in the quoted extrapolations.
\label{tabA_2-16}}
\begin{ruledtabular}
\begin{tabular}{cccccc}
 Nucleus ($J^P$)       & Exp  & Extrap (A) &  Extrap (B) & variational bound
 & Max($N_{\max}$) \\ 
\hline
$^{2}$H ($1^+$)        &  2.225 &  2.223(5) &  2.226(6) &  2.220  & 20 \\
$^{3}$H ($\frac12^+$)  &  8.482 &  8.369(1) &  8.3695(25) & 8.367 & 18 \\
$^{3}$He ($\frac12^+$) &  7.718 &  7.665(1) &  7.668(5) & 7.663 & 18  \\
$^{4}$He ($0^+$)       & 28.296 & 28.299(1) & 28.299(1) & 28.298 & 18  \\
$^{6}$He ($0^+$)       & 29.269 & 28.68(12) & 28.69(5)  & 28.473 &14 \\
$^{6}$Li ($1^+$)       & 31.995 & 31.43(12) & 31.45(5)  & 31.185 &14  \\
$^{8}$He ($0^+$)       & 31.408 & 29.74(34) & 30.05(60) & 28.927 & 12  \\
$^{12}$C ($0^+_1$)       & 92.162 & 93.9(1.1) & 95.1(2.7) & 90.9 & 8\\
$^{12}$C ($0^+_2$)       & 84.508 & 80.7(2.3) &  --- & ---  & 8\\
$^{16}$O ($0^+_1$)       &127.619 & 143.5(1.0) & 150 (14) &134.5 & 8 \\
$^{16}$O ($0^+_2$)       &121.570 & 130.6(7.6) &  --- & --- & 8\\
\end{tabular}
\end{ruledtabular}
\end{table*}

\begin{figure}
\centerline 
{\includegraphics[width=9cm]{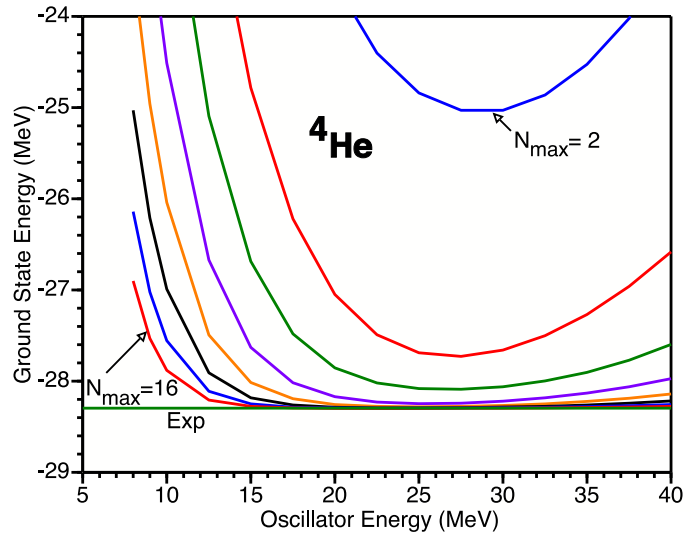}}
\caption{(Color online) Calculated ground state energy of $^4$He as
function of the oscillator energy, $\hbar\Omega$, for a sequence of
$N_{max}$ values.  The curve closest to experiment corresponds to the
value $N_{max} = 16$ and successively higher curves are obtained with
$N_{max}$ decreased by 2 units for each curve.}
\label{4He_gs_vs_hw}
\end{figure}

\begin{figure}
\centerline 
{\includegraphics[width=9cm]{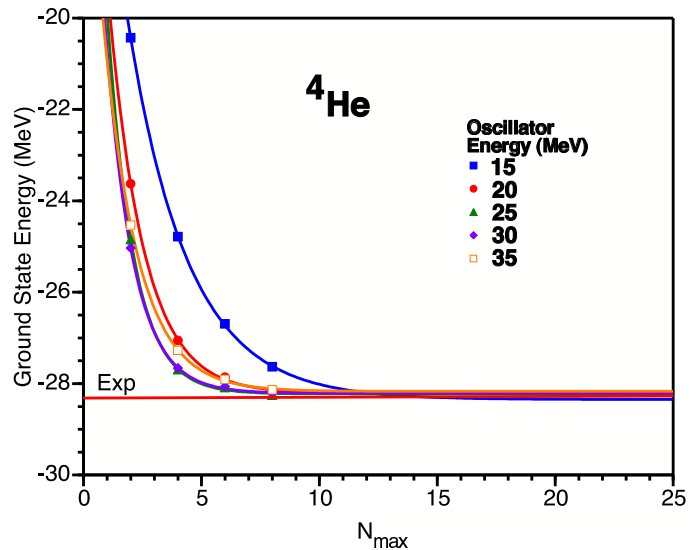}}
\caption{(Color online) Calculated ground state energy of $^4$He for
$N_{max}=2{-}8$ at selected values of $\hbar\Omega$. Each set of 4
points is fit, using equal weights, with an exponential plus constant [see
  Eq. (\ref{extreq})] 
producing the solid curves.  Half the resulting spread in the
asymptotic values is used to determine the uncertainty assigned to the
first point in Fig.~\ref{4He_gs_extrap_insert_AB} for extrapolation A 
as described in the text. }
\label{4He_gs_range2_8}
\end{figure}

\begin{figure}
\centerline 
{\includegraphics[width=9cm]{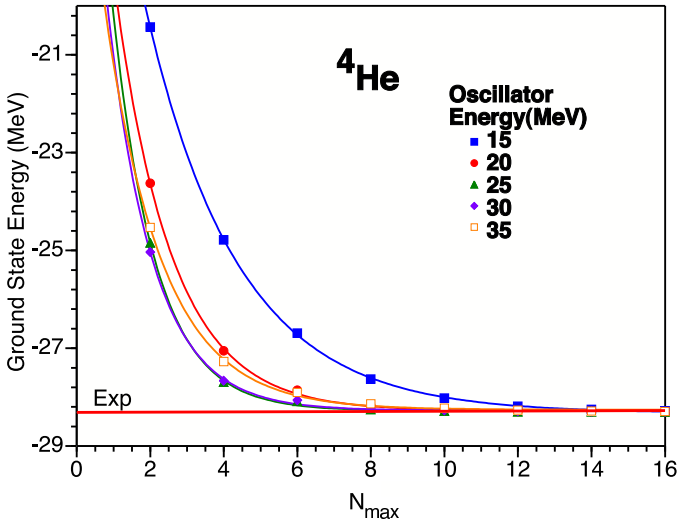}}
\caption{(Color online) Calculated ground state energy of $^4$He for
$N_{max}=2{-}16$ for JISP16 at selected values of $\hbar\Omega$.  Each
set of 8 points at fixed $\hbar\Omega$ is fit by Eq. (\ref{extreq}) 
producing the solid curves.  
Each point is a true upper bound to the exact answer.  The asymptotes
$E_{gs}({\infty})$ 
are the same to within 35~keV of their average value and they span the
experimental ground state energy.}
\label{4He_gs_range2_16}
\end{figure}

\begin{figure}
\includegraphics[width=0.99\columnwidth]{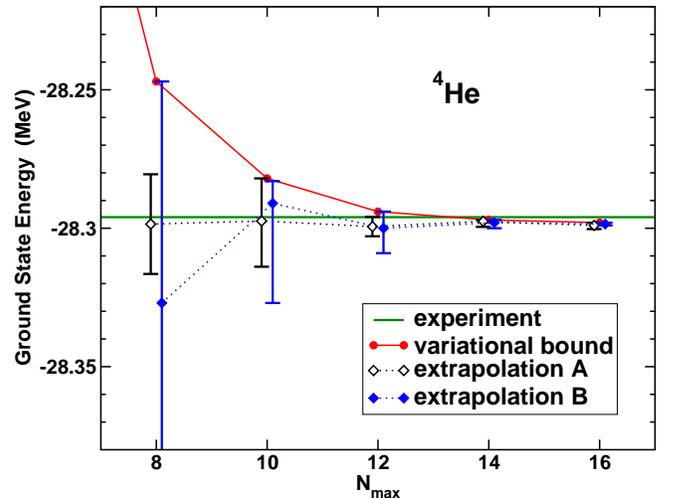}
\caption{(Color online) Extrapolated ground state energies and
variational upper bounds from each set of four (extrapolation A) or
three (extrapolation B) successive $N_{max}$ values as a function of
the largest value of $N_{max}$ in each set.  Error bars represent the assessed
uncertainties in the extrapolations and are obtained as described
in the text.  Note the consistency of the extrapolations: the exact
answer is well within the uncertainty range of the extrapolations for
all $N_{max}$ points, with the uncertainty diminishing with increasing
$N_{max}$.
\label{4He_gs_extrap_insert_AB}}
\end{figure}

As a final test of our extrapolations we consider $^4$He.  We present
our calculations as a function of $\hbar\Omega$ at fixed values
of $N_{max}$ in Fig.~\ref{4He_gs_vs_hw}.  The results clearly
indicate rapid convergence in both $N_{max}$ and $\hbar\Omega$; at
$N_{max}=16$ the ground state energy is converged to within 1 keV over
the range $20~{\rm MeV} \le \hbar\Omega \le 25~{\rm MeV}$.
Furthermore, the fully converged NCFC ground state energy is within 3
keV of the experimental energy as shown in Table~\ref{tabA_2-16}.
\begin{figure}
\includegraphics[width=0.99\columnwidth]{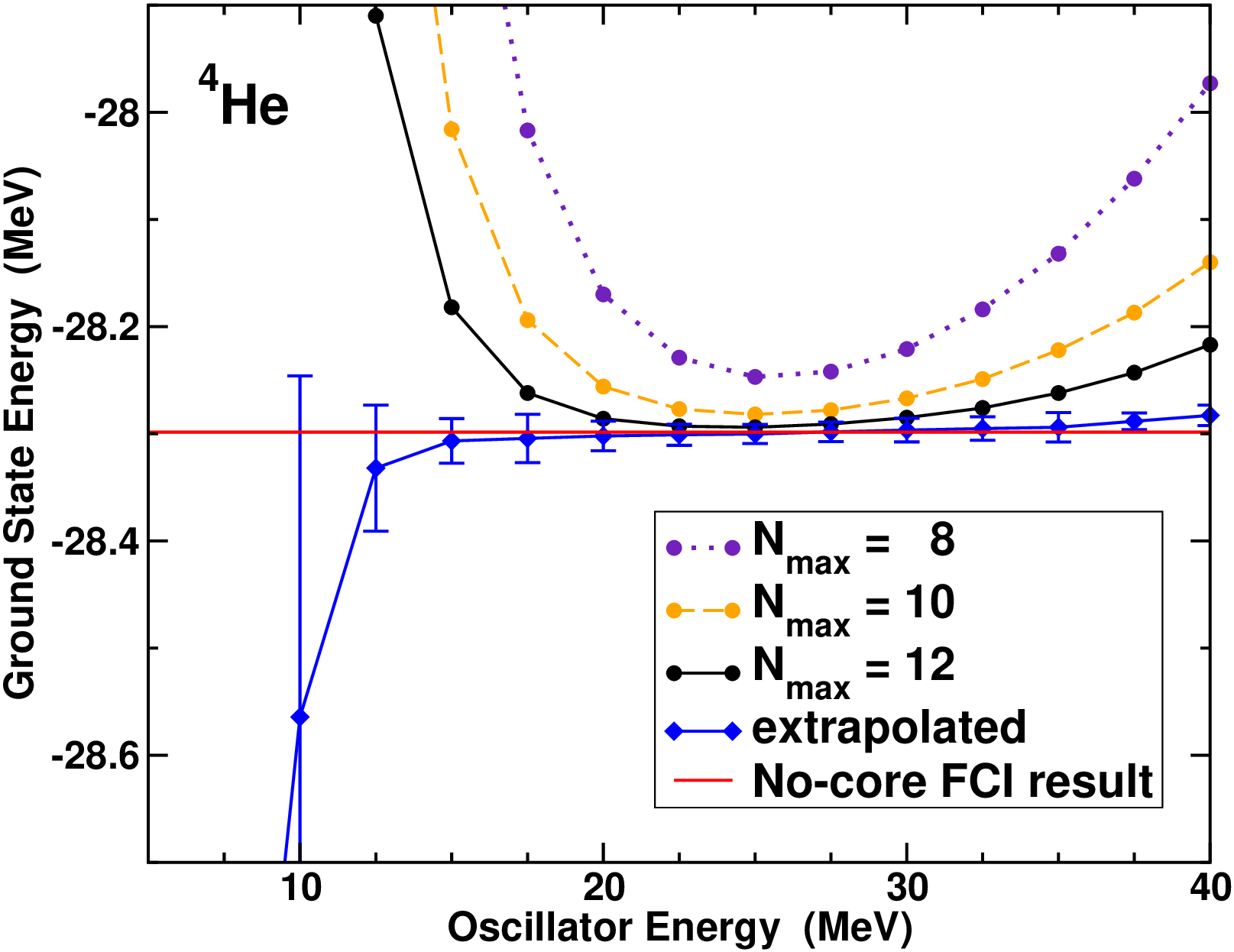}
\caption{(Color online) Calculated ground state energy of $^4$He for
$N_{max} = 8$, $10$, $12$, and the extrapolated ground state energy
using method B, as function of the oscillator energy, $\hbar\Omega$.
Error bars are obtained from the difference with the extrapolation
using $N_{max} = 6$, $8$, and $10$ calculations.
\label{4He_gs_extrapBtest}}
\end{figure}
As illustrations of our extrapolations, we demonstrate in 
Fig.~\ref{4He_gs_range2_8} the independent fits used to assess uncertainties
of extrapolation A based on $N_{max}=2$ to $N_{max}=8$ results.   In addition,
Fig.~\ref{4He_gs_extrapBtest} shows extrapolation B at fixed
values of $\hbar\Omega$ using $N_{max}=8$, $10$, and $12$ results.  

We also confirm that the dependence on $N_{max}$ at fixed
$\hbar\Omega$ is nearly a pure exponential as illustrated best in
Fig.~\ref{4He_gs_range2_16} where we show a wider range of the
calculated results.  Here, we provide regression analyses for each set
of results spanning $N_{max}=2{-}16$ at fixed $\hbar\Omega$ values
ranging from 15 to 35 MeV.  For both basis space parameters, this is a
significantly wider range of parameter values than we use in our
applications below.  

We present the NCFC results of both
extrapolation methods in
Fig.~\ref{4He_gs_extrap_insert_AB} along with the experimental and
variational upper bound energies.  In this case the results produce 
very rapid convergence
with uncertainties that drop precipitously with increasing
$N_{max}$ as seen in the figure.  We note that the error bars
conservatively represent the extrapolation uncertainties since all the
extrapolated results are, within their error bars, consistent with
each other and with the fully converged NCFC result.  The largest
$N_{max}$ points define the results quoted in Table~\ref{tabA_2-16}, a
ground state overbound by $3 \pm 1$ keV.

We have seen in this section that the NCFC results for
three light nuclei provide sufficiently converged ground state
energies to allow us to test our extrapolation methods and confirm the
validity of their assigned uncertainties.
In what follows, we present NCFC calculations for five nuclei
using both extrapolations A and B.  The five nuclei selected for this
initial application
consist of stable and unstable even nuclei that span the $p$-shell.  We
include two loosely-bound
 nuclei, $^6$He and $^8$He, anticipating that they
will provide challenges for achieving a converged ground state energy.

\subsection{NCFC results for $^6$He, $^6$Li, and $^8$He}

By comparing Figs.~\ref{2H_gs_vs_hw}, \ref{3H_gs_vs_hw} and
\ref{4He_gs_vs_hw} we observe clearly the marked correlation between
binding energy and convergence rate --- the more deeply bound ground
states exhibit greater independence of $\hbar\Omega$ at fixed
$N_{max}$.  A more complete picture of this correlation is
seen below, for example, with the $^6$He and $^6$Li results where one
observes that the relevant energy scale governing the rate of
convergence is the binding with respect to the nearest threshold.  Our
physical intuition supports this correlation since we know the
asymptotic tails of the bound state wave functions fall more slowly as
one approaches a threshold for dissociation.  This same intuition
tells us to expect Coulomb barriers and angular momenta to play
significant roles in this correlation.

\begin{figure}
\centerline 
{\includegraphics[width=9cm]{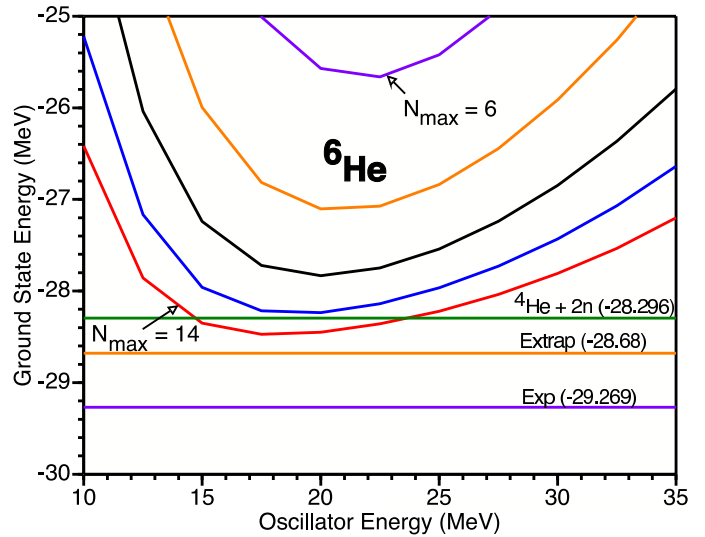}}
\caption{(Color online) Calculated ground state energy of $^6$He as
function of the oscillator energy, $\hbar\Omega$, for selected values
of $N_{max}$.  The figure displays also the experimental result, the common
asymptote from the global extrapolation (A) and the experimental
threshold for the $\alpha+2n$ breakup.
The curve closest to experiment corresponds to the value
$N_{max} = 14$ and successively higher curves are obtained with
$N_{max}$ decreased by 2 units for each curve.
\label{6He_gs_vs_hw}}
\end{figure}

\begin{figure}
\centerline 
{\includegraphics[width=9cm]{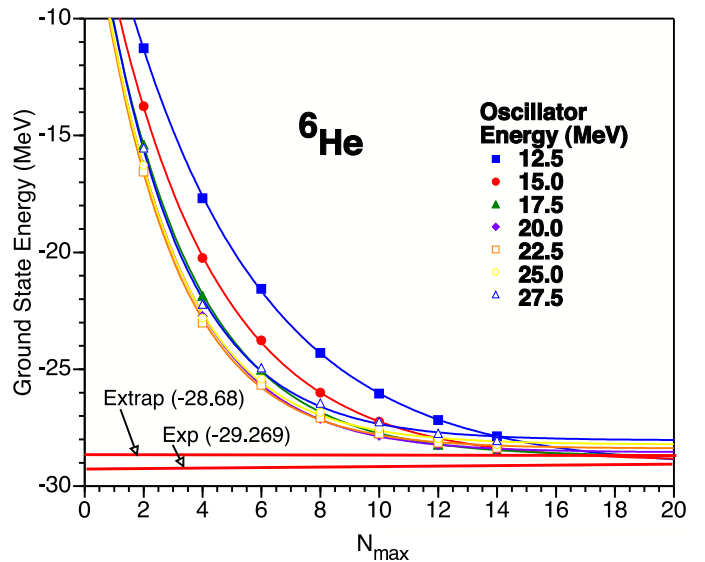}}
\caption{(Color online) Calculated ground state energy of $^6$He for
$N_{max}=2{-}14$ for JISP16 at selected values of $\hbar\Omega$.  Each
set of points at fixed $\hbar\Omega$ is fit by Eq. (\ref{extreq})
using equal weights 
producing the solid
curves.  Each point is a true upper bound to the exact answer.  The
resulting asymptotes $E_{gs}({\infty})$  are the same to within 600
keV of their average 
value.  The figure displays the experimental result and the common
asymptote from the global extrapolation (A) as described in the text.
\label{6He_gs_range2_14}}
\end{figure}

\begin{figure}
\includegraphics[width=0.99\columnwidth]{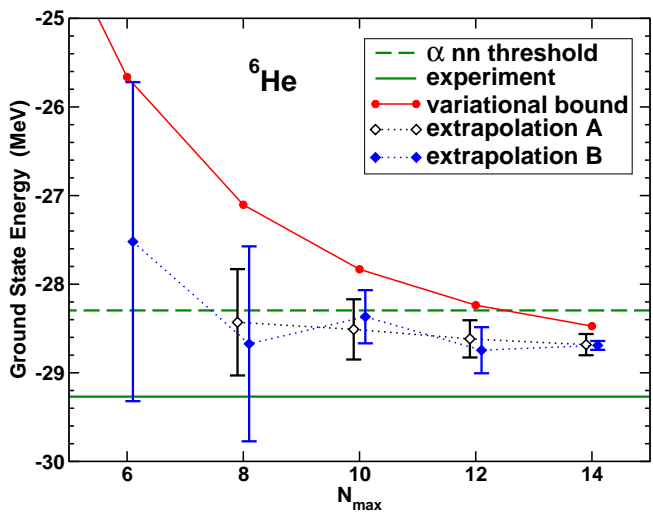}
\caption{(Color online) Extrapolated ground state energies and upper
bounds from sets of successive $N_{max}$ values as a function of the
largest value of $N_{max}$ in each set.  Uncertainties are determined
for each value of $\hbar\Omega$ as described in the text.  Note the
consistency of the extrapolated results as they fall well within their
uncertainty ranges along the paths of converging sequences.
\label{6He_gs_extrap_insert_AB}}
\end{figure}

Consider first the weakly-bound
nucleus $^6$He presented in
Figs. \ref{6He_gs_vs_hw}, \ref{6He_gs_range2_14}, and
\ref{6He_gs_extrap_insert_AB}.  In the largest basis spaces achieved,
$N_{max}=14$, 
we obtain net binding with respect
to the breakup threshold as seen in Fig.~\ref{6He_gs_vs_hw}.  However,
the results appear farther from convergence than the case for $^4$He
in the same $N_{max}$ spaces depicted in Fig.~\ref{4He_gs_vs_hw}.
Note the proximity of the $^4{\rm He}+2n$ breakup threshold to the
calculated ground state energies that suggests the importance of
achieving results at $N_{max}=16$ as well as obtaining improved
nuclear Hamiltonians that better reproduce the experimental binding of
$^6$He.

To examine this situation in some detail, we present in
Fig.~\ref{6He_gs_range2_14} results as a function of $N_{max}$ that
covers a range of 15 MeV in $\hbar\Omega$ values.  We also present the
linear regression analyses for the range of $N_{max}=2{-}14$ to show
that the exponential fit appears to maintain its validity.  Thus, we
proceed with the extrapolation methods as developed and tested in
previous sections and display the results in
Fig.~\ref{6He_gs_extrap_insert_AB}.  

Since we use only three $N_{max}$ points for our extrapolation B, at
fixed values of $\hbar\Omega$, we also include extrapolated results
based on $N_{max}=2$, $4$, and $6$; with an error estimated based on
the difference between the $N_{max}=6$ calculation and the
extrapolated result.  Such an extrapolation, based on a rather small
model space, can be useful for larger nuclei, and it turns out to be quite
reasonable.  However, we do see a rather significant 
``odd-even'' effect
with extrapolation B for this nucleus.

The results from the highest value of the upper limit in $N_{max}$ are
provided in Table~\ref{tabA_2-16}.  We again observe consistency in
the results the the global and fixed $\hbar\Omega$ extrapolations.
The JISP16 interaction yields about 600 keV underbinding in $^6$He which
implies the theoretical proton RMS radius will likely be too large
compared with experiment.

\begin{figure}
\centerline 
{\includegraphics[width=9cm]{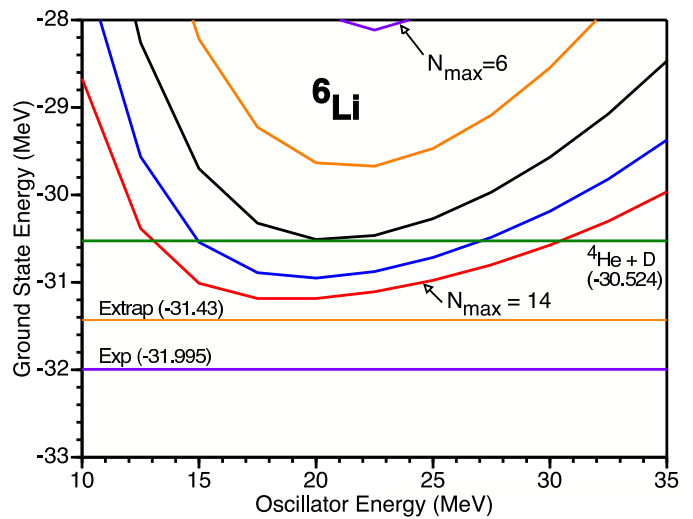}}
\caption{(Color online) Calculated ground state energy of $^6$Li as
function of the oscillator energy, $\hbar\Omega$, for selected values
of $N_{max}$. The curve closest to experiment corresponds to the value
$N_{max} = 14$ and successively higher curves are obtained with
$N_{max}$ decreased by 2 units for each curve. The figure
  displays also the global extrapolation (A) and the threshold for the 
$\alpha+d$ breakup.
\label{6Li_gs_vs_hw}}
\end{figure}

\begin{figure}
\centerline 
{\includegraphics[width=9cm]{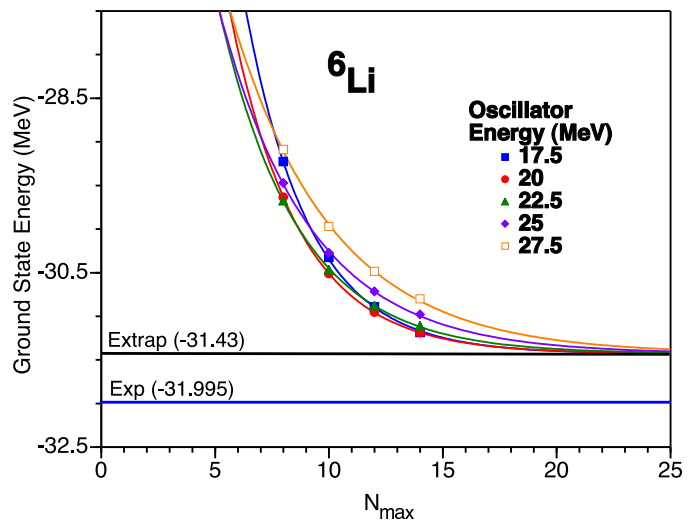}}
\caption{(Color online) Calculated ground state energy of $^6$Li for
$N_{max}=8{-}14$ at values of $\hbar\Omega$ that span the minimum at
$N_{max}=14$.  Curves define the fits using the global extrapolation
(A) that produces a common constant, the asymptote, labelled by
``Extrap".
\label{6Li_gs_extrap}}
\end{figure}

\begin{figure}
\includegraphics[width=0.99\columnwidth]{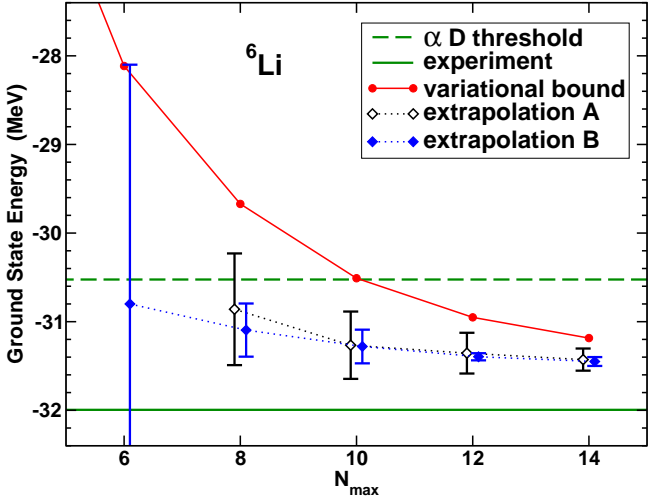}
\caption{(Color online) Upper bounds and extracted asymptotes for the
ground state energy of $^6$Li from each set of successive $N_{max}$
values as a function of the largest value of $N_{max}$ in each set.
Uncertainties in the asymptotes are determined as described in the
text.  Note the consistency of the extrapolated results as they fall
well within their uncertainty ranges along the paths of converging
sequences.
\label{6Li_gs_extrap_insert_AB}}
\end{figure}

Next, we turn our attention to $^6$Li and present our NCFC
results using extrapolations in Figs.~\ref{6Li_gs_vs_hw},
\ref{6Li_gs_extrap} and \ref{6Li_gs_extrap_insert_AB}.  As in the
$^6$He case, there is a low-lying threshold for breakup --- here about
1.47 MeV above the experimental ground state.  Both our $N_{max}$ = 12
and 14 curves drop below this threshold over a range of $\hbar\Omega$
values as seen in Fig.~\ref{6Li_gs_vs_hw}.

Our global extrapolation for $^6$Li is depicted in
Fig.~\ref{6Li_gs_extrap} where we select the case with the highest
upper limit in $N_{max}$ to portray.  The minimum in $\hbar\Omega$ at
$N_{max}=14$ occurs at $\hbar\Omega=20$ MeV.  According to our global
extrapolation, we then perform the constrained fit on the results in
the span $\hbar\Omega=17.5{-}27.5$ MeV in 2.5 MeV increments to obtain
the fits shown in Fig.~\ref{6Li_gs_extrap}.  The asymptote, the
extrapolant ($-31.43$ MeV), becomes the last data point on the right in
Fig.~\ref{6Li_gs_extrap_insert_AB} where the uncertainty is obtained
in the manner described above.  We also perform our extrapolation at
fixed values of $\hbar\Omega$, and find results consistent with our
global extrapolation, see Fig.~\ref{6Li_gs_extrap_insert_AB}.

The extrapolated results are entered in Table~\ref{tabA_2-16} and
compared with experiment.  We find that $^6$Li is underbound by about
560 keV with the JISP16 interaction, similar to the amount of
underbinging for $^6$He; the rate of convergence and error estimate in
our final answer are also similar.  However, the ``odd-even'' effect we
found with extrapolation B for the other nuclei (including $^8$He
below) is absent in $^6$Li.

\begin{figure}
\includegraphics[width=0.99\columnwidth]{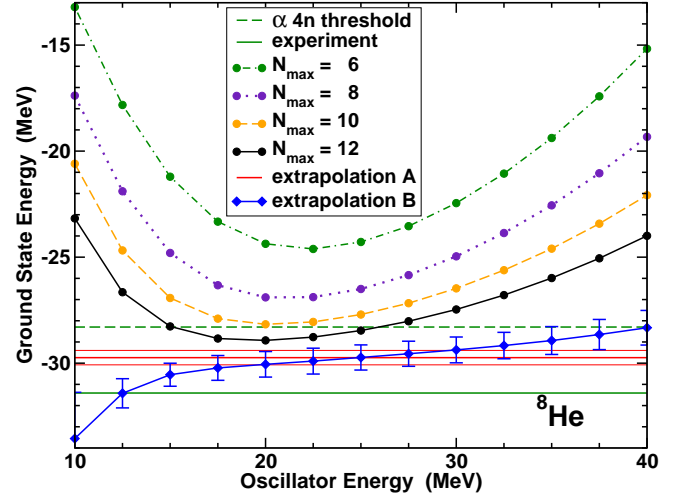}
\caption{(Color online) Calculated ground state energy of $^4$He for
$N_{max} = 6$, $8$, $10$, $12$, and the extrapolated ground state energies
using method B, as well as the result from extrapolation method A,
with error bars.  For comparison, we also show the experimental value,
and the threshold for $\alpha\, + 4n$ breakup.
\label{8He_gs_extrapBtest}}
\end{figure}

\begin{figure}
\includegraphics[width=0.99\columnwidth]{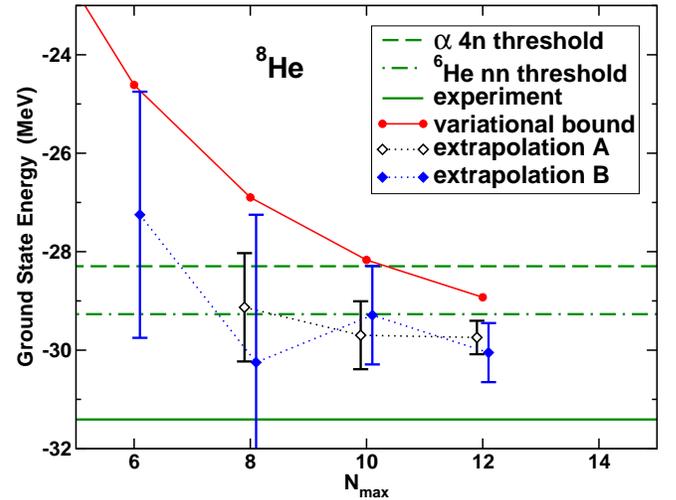}
\caption{(Color online) Upper bounds and extrapolated ground state
energies of $^8$He from each set of successive $N_{max}$ values as a
function of the largest value of $N_{max}$ in each set .
Uncertainties in the asymptotes are determined as described in the
text.\label{8He_gs_extrap_insert_AB}}
\end{figure}

Next, let us consider another  weakly-bound nucleus, $^8$He.  In
Fig.~\ref{8He_gs_extrapBtest} we show our results for the $^8$He
ground state energy as function of $\hbar\Omega$ for several values of
$N_{max}$, together with the extrapolated results at fixed
$\hbar\Omega$ and our result from the global extrapolation.  The
extrapolated energies and their uncertainties are presented in
Fig.~\ref{8He_gs_extrap_insert_AB} as function of $N_{max}$, together
with the variational upperbounds.  For comparison, we also show the
experimental value, and the thresholds for $\alpha\, + 4n$ breakup and
for $^6{\rm He}\, + 2n$ breakup.

Clearly, the results are not as well converged as those for lighter
nuclei, because we are limited to a smaller model space, $N_{max}=12$.
Nevertheless, the variational upperbound on the ground state energy is
well below the $\alpha\, + 4n$ threshold.  Furthermore, the
extrapolations are consistent with each other, and the error bars
decrease with increasing $N_{max}$.  Our final NCFC result is
not only below the $\alpha\, + 4n$ threshold but also below the
experimental $^6{\rm He}\,+ 2n$ breakup threshold, even taking into
account the uncertainty in the extrapolation, as can be seen from
Fig.~\ref{8He_gs_extrap_insert_AB} and Table~\ref{tabA_2-16}.
Compared to the experiment, we find that $^8$He is underbound by about
$1.6 \pm 0.4$~MeV with JISP16.

\section{NCFC results for $^{12}$C and $^{16}$O}

Having illustrated the application of our methods with a range of
light nuclei, we finally turn to heavier nuclei, and perform NCFC 
calculations for $^{12}$C and $^{16}$O.  For these nuclei, we can
only go to $N_{max}=8$, so we have to rely on the extrapolation
methods.

\subsection{Extrapolating $^{12}$C}

The dimension of the model space for $^{12}$C with $N_{max} = 8$ and
limited to total $M =0$ states in the $m$-scheme is 594,496,743; and the
total number of nonzero matrix elements in the lower triangle of the
extremely sparse many-body Hamiltonian matrix is 539,731,979,351 with
NN interactions only.  Thus, storage of one vector in this model space
requires 2.4 GB, and storage of the lower triangle of the matrix
requires 4.3 TB.  The dimension of the $N_{max}=10$ basis space is
7,830,355,795, which is beyond our present capabilities.  

\begin{figure}
\centerline 
{\includegraphics[width=9cm]{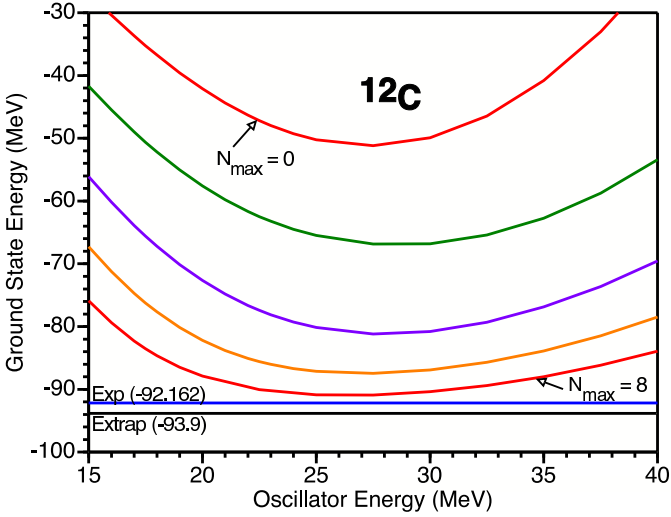}}
\caption{(Color online) Calculated ground state energy of $^{12}$C as
function of the oscillator energy, $\hbar\Omega$, for selected values
of $N_{max}$.  The curve closest to experiment corresponds to the value
$N_{max} = 8$ and successively higher curves are obtained with
$N_{max}$ decreased by 2 units for each curve.
\label{12C_gs_vs_hw}}
\end{figure}

\begin{figure}
\centerline 
{\includegraphics[width=9cm]{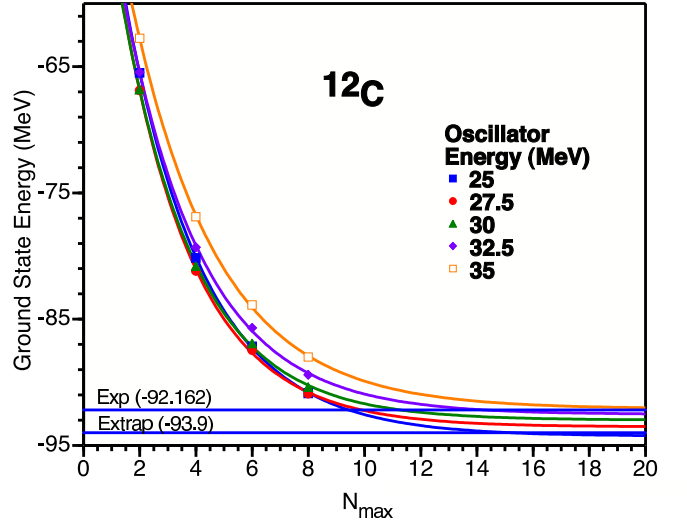}}
\caption{(Color online) Calculated ground state energy of $^{12}$C for
$N_{max}=2{-}8$ at selected values of $\hbar\Omega$ as described in the
text.  For each $\hbar\Omega$ the data are fit by Eq. (\ref{extreq}). 
These independent asymptotes  $E_{gs}({\infty})$ 
provide a measure of our uncertainty within the
global extrapolation (A). 
The figure displays the experimental ground
state energy and the common asymptote obtained in the global
extrapolation. 
\label{12C_gs_extrap}}
\end{figure}

In Fig.~\ref{12C_gs_vs_hw} we show our results for $^{12}$C
for $N_{max} = 0$ through $N_{max}=8$.  Since $N_{max} = 0$ is
generally not very reliable for our extrapolations, we only have only
the extrapolation from the $N_{max}=2{-}8$ results.  In order to
illustrate the details of our uncetainties, we depict in
Fig.~\ref{12C_gs_extrap} the linear regression analyses of our results
spanning the minimum in $\hbar\Omega$ obtained at $N_{max}=8$.  Our
global extrapolation A produces a ground state energy of $93.9 \pm
1.1$~MeV, whereas the extrapolation B at fixed $\hbar\Omega=27.5$~MeV
(where it is most stable) gives $95.1 \pm 2.7$~MeV.  Given the
``odd-even'' effect which often plagues extrapolation B, in
combination with the smaller error bar obtained with extrapolation A,
we conclude that JISP16 produces a binding energy of about $94$~MeV
with an uncertainty of 1\% to 2\%; or in other words, it overbinds
$^{12}$C by about $1.8$~MeV.

\begin{figure}
\centerline 
{\includegraphics[width=9cm]{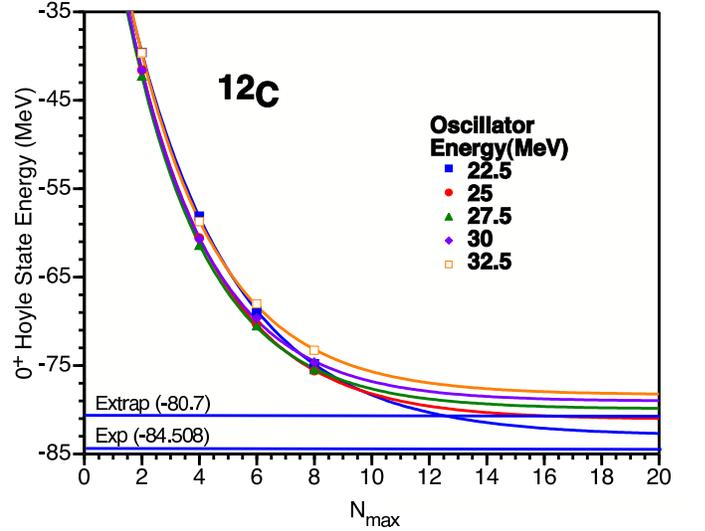}}
\caption{(Color online) Calculated energy of $^{12}$C first excited
$0^+$ state (Hoyle state) for $N_{max}=2{-}8$ at selected values of
$\hbar\Omega$ as described in the text.  For each $\hbar\Omega$ the
data are fit by Eq. (\ref{extreq}). 
These independent asymptotes  $E_{gs}({\infty})$ 
provide a measure of our uncertainty within the
global extrapolation (A).  The figure displays the experimental energy and
the common asymptote of the global extrapolation.}
\label{12C_Hoyle_Energy}
\end{figure}

For a speculative application, we also consider the first excited
$0^+$ state of $^{12}$C, the ``Hoyle state" or ``triple-alpha" state as
it has come to be known.  Since experimentally, this state, with
$E_{\rm Hoyle} = -84.51$~MeV, is just above the threshold for breakup
into three alpha's, $3\,E_{\alpha} = -84.89$~MeV, it may be poorly
converged.  On the other hand, our calculations for both
$^6$He and $^8$He at $N_{max}=2{-}8$ are above breakup into alpha plus
neutrons, but the extrapolations from these points produce results
with error bars of about 1~MeV, and agree with our best calculations
at $N_{max}=12$.  Encouraged by these results for $^6$He and $^8$He at
$N_{max}=8$, we apply the global extrapolation method A to the first
excited $0^+$ state of $^{12}$C.  The calculated results and extrapolation are
shown in Fig.~\ref{12C_Hoyle_Energy} and summarized in
Table~\ref{tabA_2-16}.  Our extrapolation gives $E_{\rm Hoyle} = -80.7
\pm 2.3$~MeV, corresponding to an excitation energy of $13 \pm 3$~MeV,
compared to an experimental excitation energy of $7.654$~MeV.  It
remains to be seen how reliable the extrapolation is for this (and
similar) states.  One may even expect this extrapolation to be 
unreliable as solutions obtained in our present, very limited, 
basis spaces may not accommodate all the essential physics 
of such excited states.  Nevertheless, assuming 
that our error estimates are realistic, our
conclusion is that JISP16 overbinds the ground state of $^12$C by an
MeV or two, but underbinds the first excited $0^+$ state by about 2 to
6 MeV.  When combined, that means it produces an excitation energy that is
significantly too large.

\subsection{Extrapolating $^{16}$O}

Finally, we consider $^{16}$O with $N_{max} = 8$.  The dimension of
the corresponding model space is 996,878,170; and the total number of
nonzero matrix elements in the lower triangle of the many-body matrix
is 805,811,591,748 with NN interactions only.  Thus, storage of one
vector in this model space requires 4.0 GB, and storage of the lower
triangle of the matrix requires 6.5 TB.

\begin{figure}
\centerline 
{\includegraphics[width=9cm]{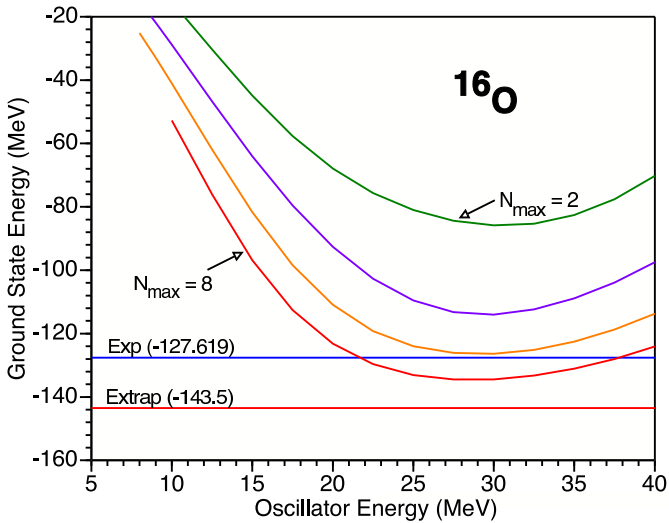}}
\caption{(Color online) Calculated ground state energy of $^{16}$O as
function of the oscillator energy, $\hbar\Omega$, for selected values
of $N_{max}$.  The curve closest to experiment corresponds to the value
$N_{max} = 8 $ and successively higher curves are obtained with
$N_{max}$ decreased by 2 units for each curve.
\label{16O_gs_vs_hw}}
\end{figure}

\begin{figure}
\centerline 
{\includegraphics[width=9cm]{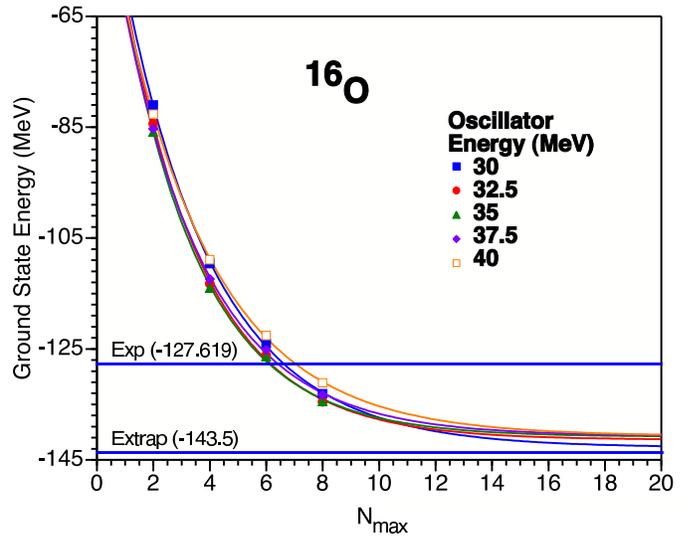}}
\caption{(Color online) Calculated ground state energy of $^{16}$O for
$N_{max}=2{-}8$ at selected values of $\hbar\Omega$ as described in the
text.  For each $\hbar\Omega$, the data are fit to an exponential plus
a constant, the asymptote.  The figure displays the experimental
ground state energy and the common asymptote obtained in the global
extrapolation (A).
\label{16O_gs_extrap}}
\end{figure}

The results for $^{16}$O are shown in Figs.~\ref{16O_gs_vs_hw} and
\ref{16O_gs_extrap}, and summarized in Table~\ref{tabA_2-16}.  As in
the case of $^{12}$C, we only attain the results through $N_{max}=8$
with our current capabilities.  Thus, we have a single extrapolant
using $N_{max}=2$, $4$, $6$, and $8$ (extrapolation A), or use
extrapolation B for $N_{max}=4$, $6$, and $8$.  The extrapolation B
gives a lower extimated ground state energy with a significantly larger
uncertainty (10\%) compared to extrapolation A.  We anticipate that
this difference is due to the ``odd-even'' effect we have seen in most
light nuclei using extrapolation B, and expect that the results
obtained with extrapolation A are more realistic for this case.  Also
note that the variational upper bound on the ground state energy is
below the experimental ground state.  That implies that JISP16
produces an overbinding of at least 7 MeV, and more likely about 15 to
18 MeV for $^{16}$O.

For another speculative application, we also consider the first
excited $0^+$ state of $^{16}$O, believed to have a significant
cluster structure.  
Experimentally, this state is very close to
threshold for breakup into $^{12}$C plus an alpha particle.
Applying extrapolation A, we find $E_{\rm excited} = -130.6 \pm
7.6$~MeV, showing an even larger extrapolation uncertainty than for the Hoyle
state in $^{12}$C.  Compared to experiment, we find an excitation
energy of $13 \pm 8$~MeV, compared to $6.05$~MeV experimentally.
Given the large uncertainty in the extrapolation for this state, we
cannot draw any conclusions without results in larger model spaces
regarding this excited state.

\begin{figure}
\centerline 
{\includegraphics[width=9cm]{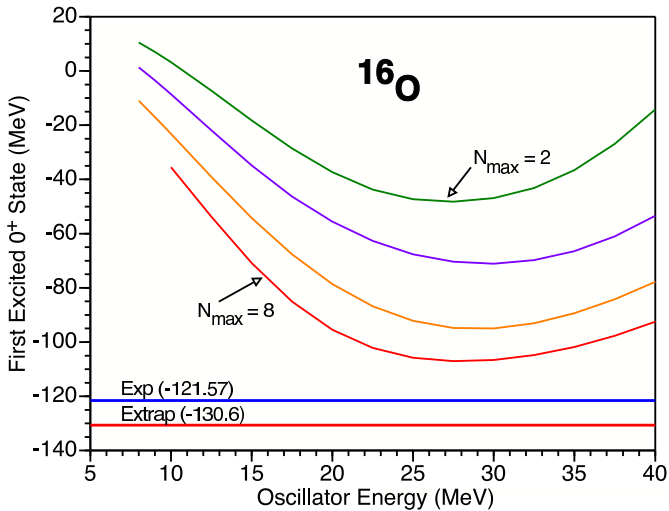}}
\caption{(Color online) Calculated first excited $0^+$ state energy of
$^{16}$O as function of the oscillator energy, $\hbar\Omega$, for
selected values of $N_{max}$.  The curve closest to experiment
corresponds to the value $N_{max} = 8 $ and successively higher curves
are obtained with $N_{max}$ decreased by 2 units for each curve.  The
figure displays the experimental excited state energy and the common
asymptote obtained in the global extrapolation (A).
\label{16O_Ex_vs_hw}}
\end{figure}

\section{Conclusions and Outlook}

We present in Table~\ref{tabA_2-16} a summary of our
results, using extrapolations, performed with methods introduced here
and compare them with the experimental results.  In all cases, we used
the calculated results to the highest $N_{max}$ available with the
bare JISP16 interaction.  In the cases of the lightest nuclei, the
extrapolations were rather modest as nearly converged results were
obtained directly.  The
uncertainties apply to the least significant digits quoted in the
table.

Our overall conclusion is that these results demonstrate sufficient
convergence is achieved for ground state energies of light nuclei to
allow extrapolations to the infinite basis limit and to estimate their
uncertainties.  Thus, we have achieve {\em ab initio} 
no-core full configuration (NCFC)
results for these nuclei with our chosen Hamiltonian.

The convergence rate reflects the short range properties of the
nuclear Hamiltonian.  Fortunately, new renormalization schemes have
been developed and applied that show promise for providing suitable
nuclear Hamiltonians with good convergence properties \cite{Bogner08}.  
Additional work is needed to develop the
corresponding NNN interaction.  Also, further work is in progress to
develop extrapolation tools for the RMS radii.  
Of course, the RMS radii present a
greater challenge since they are more sensitive than the energies to
the asymptotic properties of the wavefunctions.

\section{Acknowledgments}

We thank Miles Aronnax for calculating some of the dimensions plotted in
Fig.~\ref{Dimensions}.  We thank Petr Navr\'atil for the chiral N3LO
results shown in Fig.~\ref{4He_JISP16_v_N3LO}.  We also thank Richard
Furnstahl, Petr Navr\'atil and Christian Forssen for useful
discussions.  This work was supported in part by the U. S. Department
of Energy Grants DE-FC02-07ER41457
 and DE-FG02-87ER40371 and by the
Russian Foundation of Basic Research.   Results are
obtained through grants of supercomputer time at NERSC and at ORNL.
The ORNL resources are obtained under the auspices of an INCITE award
(David Dean, PI).  We especially wish to acknowledge MFDn code
improvements \cite {Sternberg08} developed in collaboration with Masha Sosonkina (Ames
Laboratory), Hung Viet Le (Ames Laboratory), Anurag Sharda (Iowa State
University), Esmond Ng (LBNL), Chao Yang (LBNL) and Philip Sternberg
(LBNL). 



\begin{thebibliography}{10}
\bibitem[*]{USG} This report was prepared as an account of work sponsored
by an agency of the United States Government. Neither the United States
Government nor any agency thereof, nor any of its employees, makes any
warranty, express or implied, or assumes any legal liability
or responsibility for the accuracy, completeness, or
usefulness of any information, apparatus, product, or process
disclosed, or represents that its use would not infringe
privately owned rights.  Reference herein to any  
specific commercial product, process, or service by trade name, trademark, 
manufacturer, or otherwise does not necessarily constitute or imply its 
endorsement, recommendation, or favoring by the United States Government
or any agency thereof. The views and opinions of the authors expressed herein
do not necessarily state or reflect those of the United States Government
or any agency thereof.
\bibitem{bare} By ``bare'', we mean that we have not performed renormalization by means of
a Lee--Suzuki--Okamoto or other effective interaction approach.
\end{thebibliography}
\end{document}